\def \Et {$E_{T}$ }
\def \Pt {$p_{T}$ }

\documentstyle[12pt,epsf]{article}
\begin{document}
\baselineskip 18pt
\begin{flushright}
FERMILAB-PUB-95/333\\
October 5, 1995
\end{flushright}
\begin{center}
\vspace{1cm}
\large{
\bf{The Analysis of Multijet Events Produced at High Energy Hadron
Colliders}
}
\end{center}
\vspace{.5cm}
\begin{center}
\large{S. Geer} \\
\vspace{.4cm}
\it{Fermi National Accelerator Laboratory \\
P.O. Box 500, Batavia, Illinois 60510} \\
\vspace{1.0cm}
\large{T. Asakawa} \\
\vspace{.4cm}
\it{University of Tsukuba}\\
Tsukuba, Ibaraki 305, Japan
\end{center}
\vspace{1cm}
\normalsize

\vspace{0.4cm}

\large
\normalsize

\vspace{0.6cm}
\begin{center}
\large{Abstract}
\end{center}
\vspace{0.5cm}
We define and discuss a set of (4N - 4) parameters that can be used to
analyse events in which N jets have been produced in high energy
hadron-hadron
collisions. These multijet variables are the multijet mass and (4N - 5)
independent dimensionless parameters. To illustrate the use of the variables
QCD predictions are presented for events with up to five jets
produced at the Fermilab Tevatron Proton-Antiproton Collider.
These QCD predictions are compared with the predictions
of a model in which multijet events uniformly populate the N-body phase-space.
\\\\\\
Submitted to Physical Review D.
\newpage

\begin{section} {Introduction}

Large samples of events containing two or more jets have recently
been recorded at the Fermilab Tevatron Proton-Antiproton Collider.
Many of the observed events contain three-, four-, or even five-or-more jets
\cite{multijet_paper}.
A comprehensive analysis of these multijet events would provide an interesting
test of perturbative Quantum Chromodynamics (QCD). In addition,
a detailed understanding of the properties of multijet events produced in
high energy hadron-hadron collisions
is important, firstly because multijet production is expected to be prolific
in future high luminosity running at the Fermilab Proton-Antiproton
Collider and at the Large Hadron Collider at CERN, and secondly
because an understanding of QCD multijet production is required to facilitate
the search for more exotic processes producing multijet events. For example,
a detailed understanding of the properties
of six-jet events at the Fermilab collider is likely to be important
in the near future for
the study of $t\overline{t}$ production and decay in the
all hadronic channel.

In the past, elegant analyses of two-jet and three-jet
production have been published by the UA1 \cite{ua1_2jet,ua1_3jet}
and UA2 \cite{ua2_2jet,ua2_3jet} collaborations at the CERN
$S\overline{p}pS$ Collider and by the CDF \cite{cdf_2jet,cdf_3jet}
and D0 \cite{d0_34jet} collaborations at the
Fermilab Tevatron Collider.
There have also been analyses of events with more than
three jets \cite{d0_34jet,ua2_4jet,cdf_4jet}.
However, the analyses of events with
four or more jets have not used a simple set of independent
variables that (i) span the multijet parameter space, (ii) make it simple to
interpret the observed event distributions within
the framework of perturbative QCD, and (iii) make it is easy to compare the
characteristics of events having
N jets with the characteristics of events having for example (N+1) jets.
In this paper we discuss a set of multijet parameters that satisfy these
criteria.

In choosing a set of multijet variables that span the multijet parameter space
it should be noted that we can completely define a system of N
particles in the N-body rest-frame
by specifying 4N independent parameters, for example the
4N components of four-momentum. The N-body system would then be overspecified
since momentum conservation provides us with three constraints.
Furthermore, we can rotate
the N-body system about the incoming beam direction without loosing any
interesting information. Therefore, to describe the system we need
only specify (4N~-~4) parameters. We will take these parameters
to be the N-body mass and (4N~-~5) additional variables.
We therefore introduce and discuss a set (4N~-~5)
dimensionless variables which, with the addition of the multijet mass,
span the multijet parameter space.
Our (4N~-~5) multijet variables will
provide a simple framework within which the properties of multijet events
can be compared with QCD predictions.
To illustrate the use of these variables we will compare predictions for the
population of events in the multijet parameter space obtained
from QCD matrix element calculations with
the corresponding predictions from
QCD parton shower Monte Carlo calculations,
and from a model in which the events are uniformly distributed over
the available phase-space. The QCD and phase-space calculations are described
in section 2. In section 3 the analysis of two-jet events is briefly
discussed. The standard three-jet variables are reviewed and extended
in section 4. Four-jet
and five-jet variables are introduced and discussed in sections 5 and 6. In
section 7 the generalization of the multijet
parameters to describe topologies with more than five jets is discussed.
Finally, a summary is given in section 8.

\end{section}

\begin{figure}
\epsfysize=3.0in
\vspace{0.5cm}
\epsffile[-120 144 522 590]{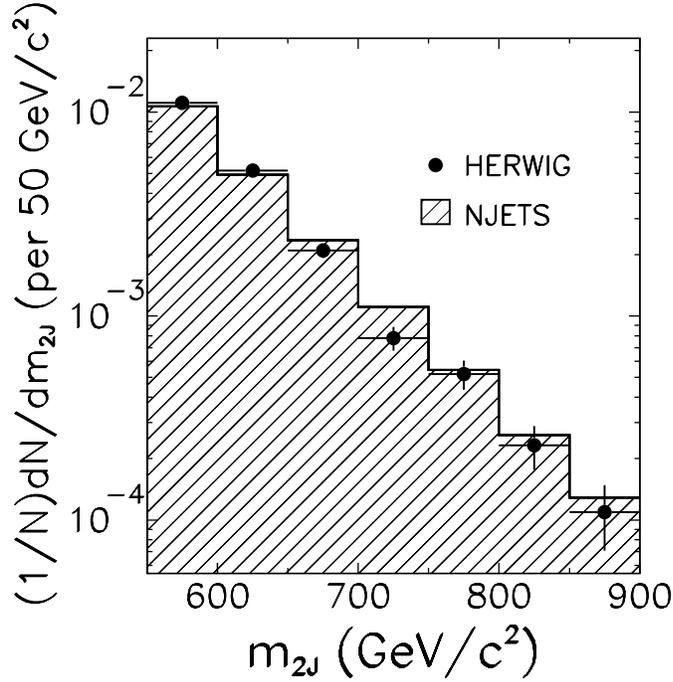}
\vspace{-0.1cm}
\caption{Predicted two-jet mass distributions for two-jet events
produced at the Fermilab Proton-Antiproton Collider. HERWIG (points)
compared with NJETS (histogram) after applying the requirements
of $m_{2J}>550$ GeV/$c^2$ and $\mid\cos\theta^\star\mid <$ 0.6.}
\label{m2j_fig}
\end{figure}

\begin{figure}
\epsfysize=3.0in
\vspace{0.8cm}
\epsffile[-120 144 522 590]{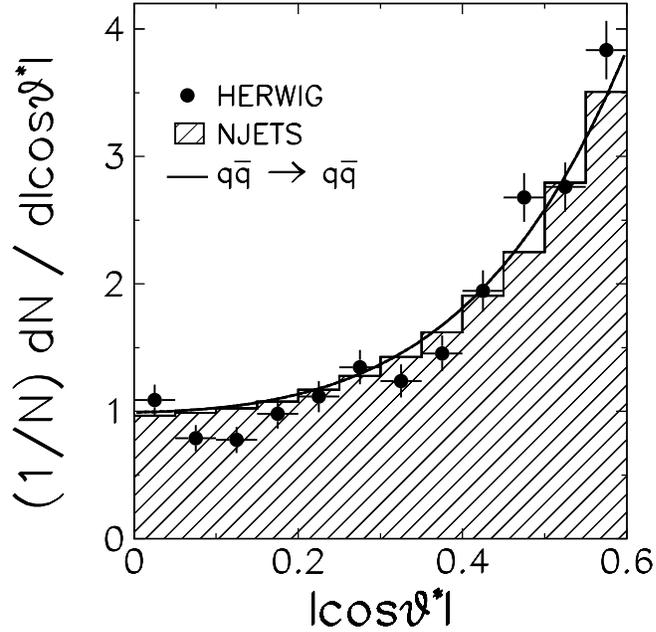}
\vspace{-0.2cm}
\caption{Predicted $\mid\cos\theta^\star\mid$ distributions for two-jet
events produced at the Fermilab Proton-Antiproton Collider that satisfy
the requirements
$m_{2J}> 550$ GeV/$c^{2}$ and $\mid\cos\theta^\star\mid<0.6$.
The HERWIG prediction (points) is
compared with the NJETS prediction (histogram), and
the LO QCD prediction for $q\overline{q} \rightarrow q\overline{q}$ scattering
(curve).}
  \label{cos2j_fig}
\end{figure}

\begin{figure}
\epsfysize=5.0in
\vspace{0.5cm}
\epsffile[30 144 522 590]{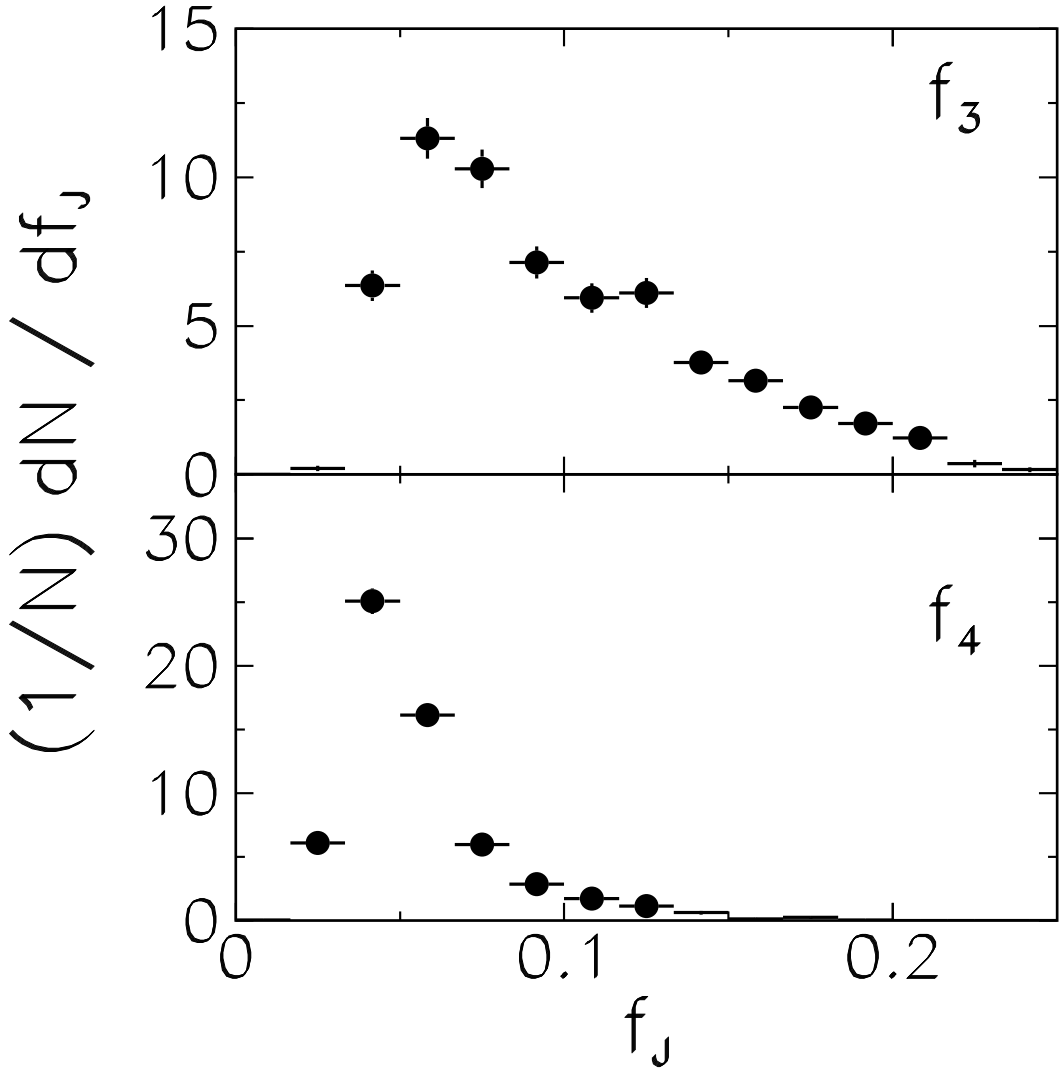}
\caption{The HERWIG Monte Carlo predictions for the
distributions of leading and next-to-leading single-jet-mass
fractions for jets in two-jet events produced at the Fermilab
Proton-Antiproton Collider that satisfy
the requirements $m_{2J}> 550$ GeV/$c^{2}$ and
$\mid\cos\theta^\star\mid < 0.6$.}
  \label{fj_2j_fig}
\end{figure}

\begin{figure}
\epsfysize=3.0in
\vspace{0.5cm}
\epsffile[-120 144 522 590]{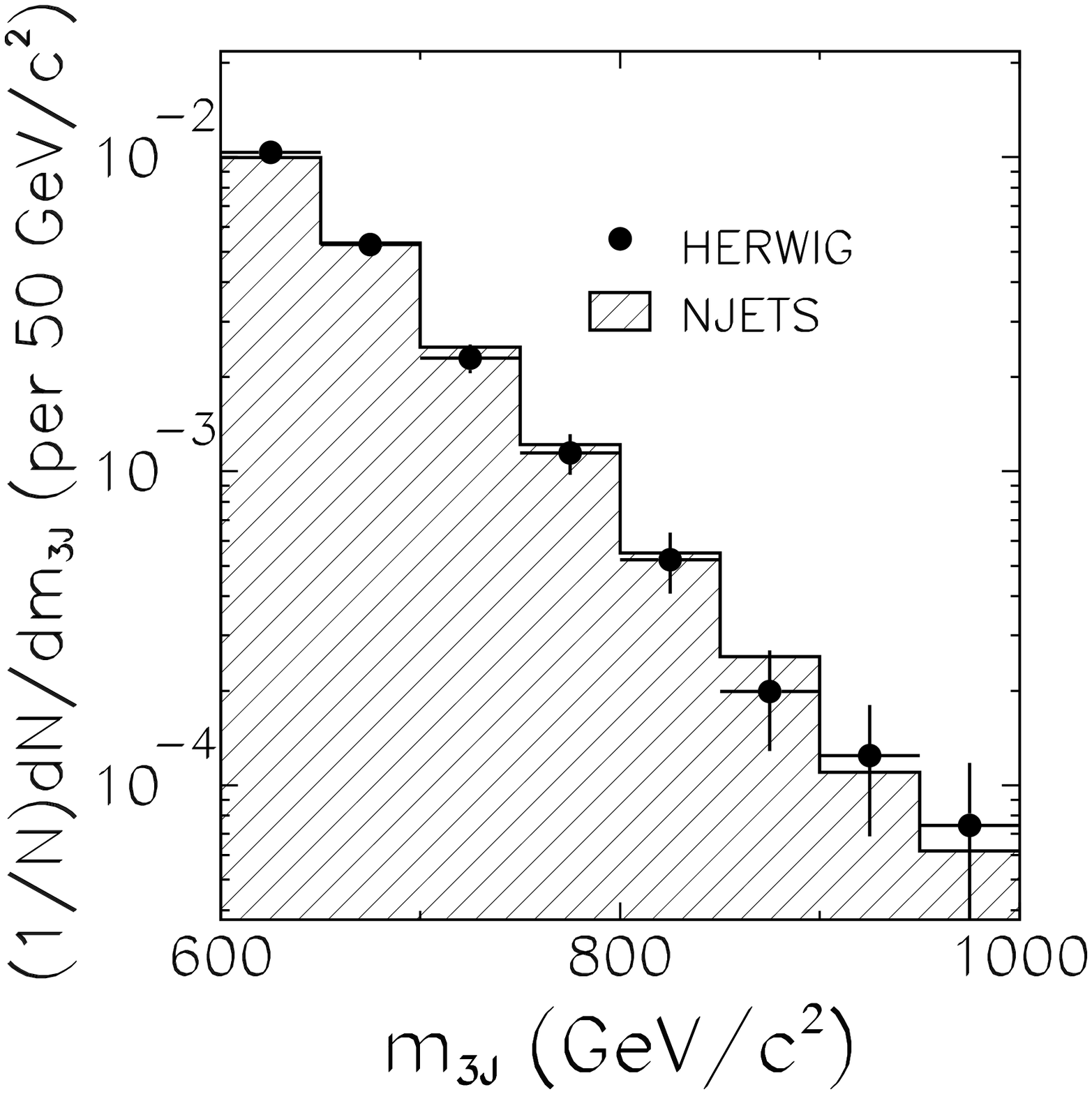}
 \caption{Predicted three-jet mass distributions for events
produced at the Fermilab Proton-Antiproton Collider that satisfy the
requirements $m_{3J}>600$ GeV/$c^{2}$, $X_3<0.9$, and
$\mid \cos\theta_{3}\mid < 0.6$.
HERWIG predictions (points) are compared with
NJETS predictions (histogram).}
  \label{m3j_fig}
\end{figure}

\begin{figure}
\vspace{-0.5cm}
 \epsfysize=4.0in
 \epsffile[ -63 210 522 648]{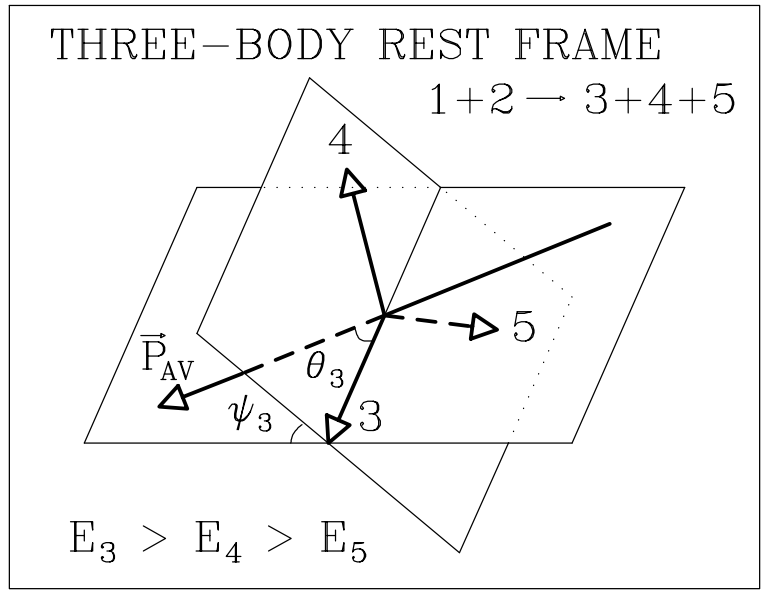}
 \vspace{-3.5cm}
\vspace{1.0in}
 \caption{Schematic definition of angles used to describe the three-jet
system in the three-jet rest-frame.}
 \label{3jet_angle_fig}
\end{figure}

\begin{section} {QCD and Phase-Space Predictions}

To illustrate the use of our multijet variables we will present and discuss
various predictions for the distribution of multijet events in the multijet
parameter space. In  particular we will consider
two-jet, three-jet, four-jet, and five-jet events produced at the Fermilab
Proton-Antiproton Collider operating at a center of mass energy of 1.8 TeV,
and compare predictions obtained from:
(a) the HERWIG {\cite{HERWIG}} QCD parton shower Monte Carlo program,
(b) the NJETS {\cite{NJETS}} Leading Order (LO) QCD
$2 \rightarrow N$ matrix element Monte Carlo program, and
(c) a model in which events are distributed uniformly over the available
N-body phase-space.

\begin{subsection} {Jet definitions and selection criteria}

The QCD and phase-space model predictions depend upon the
algorithm used to define jets and the selection criteria
used to define the data sample.
To illustrate the use of our multijet variables we will take as an
example jet definitions and event selection criteria recently used by the
CDF collaboration to define a multijet data sample recorded
at the Fermilab Proton-Antiproton Collider \cite{multijet_paper}.
Our predictions will therefore be for an existing data sample.
Following the CDF prescription,
jets are defined such that they satisfy the following:
\begin{description}
\item [{(i)}]  Jet transverse energy \Et $> 20$ GeV,
where \Et $\equiv E \sin\theta$, E is the jet energy, and
$\theta$ is the angle between the jet and the beam direction in the
laboratory frame,
\item [{(ii)}] $\mid \eta \mid < 3$,
where the jet pseudorapidity $\eta \equiv -\log\tan(\theta/2)$, and
\item [{(iii)}] jet-jet separation
$\triangle R > 0.9$, where
$\triangle R \equiv (\triangle\eta^2 + \triangle\phi^2)^{1/2}$, and
$\triangle\eta$ and $\triangle\phi$ are the differences in pseudorapidity
and azimuthal angle between the two jets.
\end{description}

With these jet definitions, the multijet event sample is defined by
selecting events that satisfy the following:
\begin{description}
\item [{(a)}] Total transverse energy
$\sum E_{T} > 420$ GeV,
where the sum is over all jets with $E_T >$ 20 GeV,
\item [(b)] Multijet mass $m > m_{min}$, and
\item [(c)] the cosine of the leading-jet scattering angle
$\cos\theta < (\cos\theta)_{max}$
where the leading-jet is defined as the highest energy jet in the
multijet rest-frame.
\end{description}
Note that for two-jet events the $\sum E_{T}$ requirement selects events
with jet $E_T > 210$~GeV. At fixed two-jet mass this results in an
effective maximum allowed value of $\cos\theta$. The values of $m_{min}$
and $(\cos\theta)_{max}$ are chosen to restrict the parameter
space to the region in which the $\sum E_{T}$ requirement is efficient.
%
\end{subsection}
\begin{subsection} {The HERWIG parton shower Monte Carlo calculation}

HERWIG {\cite{HERWIG}} is a leading-order QCD
parton shower Monte Carlo program that includes both
initial- and final-state gluon radiation.
HERWIG predictions can be thought of as LO QCD
$2 \rightarrow 2$ predictions with gluon radiation, color coherence,
hadronization, and an underlying event.
We have used version 5.6 of the HERWIG Monte Carlo
program together with a simple detector simulation that modifies the
jet energies with a Gaussian resolution function:
\begin{equation}
 \sigma_{E} = 0.1 \; E \; .
\end{equation}
This is similar to the jet energy resolution function reported by the
CDF collaboration \cite{multijet_paper}.
In our HERWIG calculations we have used the CTEQ1M~\cite{cteq}
structure functions and the scale Q$^2$ = stu/2(s$^2$+u$^2$+t$^2$).
HERWIG generates $2 \rightarrow 2$
processes above a specified $p_T^{hard}$ where $p_T^{hard}$ is the \Pt of
the outgoing partons from the hard scatter before any radiation has occurred.
We have set the minimum $p_T^{hard}$ to 60 GeV/c.
Finally, the HERWIG Monte Carlo distributions discussed in this paper are
inclusive. Hence, for a given jet multiplicity N, the generated
events contribute to the distributions if they have at least N jets that
pass the jet requirements. If there are more than N jets in a generated event,
the multijet system is defined using the N highest $E_T$ jets.

\end{subsection}
\begin{subsection} {The NJETS QCD matrix element calculation}

The NJETS Monte Carlo program \cite{NJETS}
provides parton-level predictions based on the
LO QCD $2 \rightarrow N$ matrix elements. We have used the KMRSD- structure
function parameterization's~\cite{kmrsd} with the renormalization scale
chosen to be the average $p_T$ of the outgoing partons.
NJETS does not use a parton fragmentation model.
Jet definitions and selection cuts are therefore applied to the
final state partons.
To enable a direct comparison between NJETS and HERWIG predictions we have
smeared the final state parton energies in our NJETS calculations
with the jet energy resolution function described above.

In the following we will find that the NJETS and
HERWIG predictions are generally in good agreement with one another.
This suggests that the QCD predictions for the distributions discussed in this
paper are probably not sensitive to reasonable variations in the choice
of structure functions, $Q^2$ scale,
jet fragmentation (with the exception of the single-jet mass distributions),
or underlying event.

\end{subsection}
\begin{subsection} {Phase-Space model}

We have generated samples of Monte Carlo events for which the
multijet systems uniformly populate the N-body phase-space.
These phase-space Monte Carlo
events were generated with single-jet masses distributed according to the
single-jet mass distribution predicted
by the HERWIG Monte Carlo program. In addition, the multijet mass distributions
were generated according to the corresponding distributions obtained
from the HERWIG Monte Carlo calculation.
Comparisons between the resulting phase-space model distributions and
the corresponding HERWIG and NJETS Monte Carlo distributions help us to
understand which multijet parameters are most sensitive to the
behaviour of QCD multijet matrix elements.

\end{subsection}
\end{section}
\begin{section} {Two-Jet Variables}

We begin by briefly reviewing the variables that are
often used in two-jet analyses \cite{ua1_2jet, ua2_2jet, cdf_2jet}.
Consider a system of two massless jets. The massless jet approximation is
appropriate because at high center-of-mass energies single-jet masses
are much smaller than two-jet masses ($m_{2J}$).
To describe a system of two massless jets in the two-jet
rest-frame we need only two variables.
In previous two-jet analyses these variables have often been chosen to be
$m_{2J}$ and $\cos \theta^\star$,
where $\theta^\star$ is the scattering angle between the incoming beam
particles and the outgoing jets in the two-jet rest-frame. In defining
$\cos \theta^\star$ it must be remembered that in practice
a two-jet system will always be produced together with a spectator system, and
the incoming beam particles will not be collinear in the two-body rest-frame.
Hence, following the convention of Collins and Soper \cite{collins_soper}
$\theta^\star$ is taken to be the angle between the outgoing jets and
the average beam direction. Consider the process $1 + 2 \rightarrow 3 + 4$.
The center-of-mass scattering angle is defined:
\begin{equation}
  \cos\theta^{\star}  \; \equiv \; \frac{\overrightarrow{P}_{AV}\cdot
                      \overrightarrow{P}_{3}}
                   {\mid \overrightarrow{P}_{AV}\mid
             \mid \overrightarrow{P}_{3} \mid} \; , \\
\end{equation}
where
\begin{eqnarray}
\overrightarrow{P}_{AV} & = & \overrightarrow{P}_1 - \overrightarrow{P}_2 \; ,
\end{eqnarray}
and we define particle 1 as the incoming interacting parton with the highest
energy in the laboratory frame.

NJETS and HERWIG QCD Monte Carlo predictions for the $m_{2J}$ and
$\cos\theta^{\star}$
distributions are shown in Figs.~\ref{m2j_fig} and \ref{cos2j_fig}
respectively for two-jet events produced at the Fermilab
Proton-Antiproton Collider satisfying the requirements  $m_{2J} > 550\;GeV/c^2$
and $\mid\cos\theta^{\star}\mid \;< 0.6$.
Note that in the HERWIG Monte Carlo calculation the jets acquire mass in the
fragmentation process, whereas in the NJETS calculation jets are identified
with massless partons. Hence the agreement between the HERWIG and NJETS
predictions reflects the validity of the massless jet approximation.
The predicted $\cos\theta^{\star}$ distributions are similar to
the angular distribution expected at LO for
$q\overline{q} \rightarrow q\overline{q}$ scattering
\cite{combridge}, which
is not very different from the well known Rutherford scattering
form:
\begin{eqnarray}
\frac{d\sigma}{d\cos\theta^{\star}} & \simeq & (1-\cos\theta^{\star})^{-2} \; .
\end{eqnarray}
Hence, the $\cos\theta^\star$ variable has some nice features.
Firstly, the LO QCD prediction for the $\cos\theta^\star$ distribution
is well known and is similar, although not identical, to the
Rutherford scattering distribution. Secondly, the
phase-space density is independent of $\cos\theta^\star$. Therefore the
measured $\cos\theta^\star$ distribution depends upon the
underlying $2 \rightarrow 2$ matrix element in a very direct way.

To prepare for the analysis of systems with many jets in
the final state it is useful to
extend the two-jet variables to describe two-jet systems with massive final
state jets. To do this we must specify two additional parameters.
Obvious choices are the final state single-jet masses $m_3$ and $m_4$.
We prefer to use dimensionless variables, and therefore define:
\begin{eqnarray}
f_3 & \equiv & \frac{m_3}{m_{2J}}\; ,
\end{eqnarray}
and
\begin{eqnarray}
f_4 & \equiv & \frac{m_4}{m_{2J}}\; ,
\end{eqnarray}
where the jets are ordered in the two-body rest-frame
such that $E_3 > E_4$, and hence $f_3 > f_4$.
The HERWIG predictions for the $f_3$ and $f_4$ distributions are shown in
Fig.~\ref{fj_2j_fig}. Note that $f_3$ and $f_4$ are typically of order 0.05
to 0.1,
and hence single-jet masses can be neglected for many purposes.

We conclude by noting that we have defined four variables
that specify a two-jet system in the two-body rest-frame:
$m_{2J}$, $\cos\theta^{\star}$, $f_3$, and $f_4$.
\end{section}
\begin{section} {Three-Jet Variables}

In the standard three-jet analysis used by the UA1 collaboration
\cite{ua1_3jet},
and later by the CDF \cite{cdf_3jet}
and D0 \cite{d0_34jet} collaborations, five variables are chosen that
specify the system of 3 massless particles in the three-body
rest-frame. The first of these variables is the three-jet mass ($m_{3J}$).
The NJETS and HERWIG predictions for the $m_{3J}$ distribution are
shown in Fig.~\ref{m3j_fig} to be in good agreement with each other. The
predicted $m_{3J}$ distributions have also recently been shown to be in
good agreement with the observed CDF $m_{3J}$ distribution
\cite{multijet_paper}.
To complete the description of the three-jet system four additional
dimensionless variables are defined
that, together with $m_{3J}$, span the three-body parameter space.
In defining the three-jet parameters
it is traditional to label the outgoing jets 3, 4, and 5, and order
the jets such that $E_3 > E_4 > E_5$, where $E_j$ is the energy of jet $j$
in the three-jet rest-frame. The traditional three-jet variables employed
are $X_3$,  $X_4$, $\cos\theta_3$, and
$\psi_3$,
which are defined:
\begin{itemize}
\item[(i)] $X_3$, the leading-jet energy fraction, normalized:
\begin{equation}
X_3 \equiv  \frac{2\;E_3}{E_3+E_4+E_5}  = \frac{2\;E_3}{m_{3J}} \; ,
\end{equation}
\item[(ii)] $X_4$, the next-to-leading jet energy fraction, normalized:
\begin{equation}
X_4  \equiv  \frac{2\;E_4}{E_3+E_4+E_5} = \frac{2\;E_4}{m_{3J}} \; ,
\end{equation}
\item[(iii)] $\cos\theta_3$, defined in the three-jet rest-frame as
the cosine of the leading-jet scattering angle
(see Fig.~\ref{3jet_angle_fig}) :
\begin{equation}
  \cos\theta_{3}  \equiv  \frac{\overrightarrow{P}_{AV}\cdot
                      \overrightarrow{P}_{3}}
                   {\mid \overrightarrow{P}_{AV}\mid
             \mid \overrightarrow{P}_{3} \mid} \; . \\
\end{equation}
\item[(iv)] $\psi_3$, defined in the three-jet rest-frame as
the angle between the three-jet plane and the
plane containing jet 3 (the leading jet) and the average beam direction
(see Fig.~\ref{3jet_angle_fig}) :
\begin{eqnarray}
  \cos\psi_{3}  & \equiv &
\frac{{(\overrightarrow{P}_{3} \times
      \overrightarrow{P}_{AV})} \cdot
      {(\overrightarrow{P}_{4} \times \overrightarrow{P}_{5})}}
{\mid {\overrightarrow{P}_{3} \times \overrightarrow{P}_{AV}}\mid
\mid {\overrightarrow{P}_{4} \times \overrightarrow{P}_{5}}\mid} \; .
\end{eqnarray}
\end{itemize}

Predictions for the $X_3$--,
$X_4$--, $\cos\theta_3$--, and $\psi_3$--distributions are shown in
Fig.~\ref{3j_three_fig} for three-jet events produced at the Fermilab
Proton-Antiproton Collider that satisfy the requirements
$m_{3J}>600$ GeV/$c^{2}$, $\mid \cos\theta_{3}\mid < 0.6$,
and $X_3<0.9$. These selection criteria are used to restrict the parameter
space to the region for which the $\sum E_{T}$ requirement is efficient and to
ensure that the jets in the three-jet sample are well measured.
The first and second three-jet parameters ($X_3$ and $X_4$) are Dalitz
variables, normalized so that $X_3+X_4+X_5 = 2$. Momentum conservation
restricts the ranges of the Dalitz variables (for massless jets
$2/3 \leq X_3 \leq 1$
and $1/2 \leq X_4 \leq 1$). The phase-space density is uniform over the
kinematically allowed region of the ($X_3,X_4$)-plane, and hence the
phase-space model predictions for the $X_3$-- and $X_4$--distributions can
be easily understood.
Note that the QCD predictions for the $X_3$-- and $X_4$--distributions are
similar to those of the phase-space model. We might have expected
the QCD calculations to predict an enhanced event rate as $X_3 \rightarrow 1$
and the three-jet system therefore approaches a two-jet configuration.
However, in practice the algorithm used to define jets and the experimental
requirements used to select well measured three-jet
events restrict the measured three-jet topologies to those that populate
regions of the three-body phase-space where the matrix element varies only
slowly over the ($X_3,X_4$)-plane.
The third and fourth three-jet parameters
($\cos\theta_3$ and $\psi_3$) are angular variables. The phase-space
density is uniform in $\cos\theta_3$-space, $\psi_3$-space, and
is also uniform in the ($\cos\theta_3$, $\psi_3$)-plane. Indeed,
the phase-space model does predict a uniform $\cos\theta_3$ distribution.
The phase-space model prediction for the $\psi_3$ distribution is not quite
uniform, there being a slight depletion of events as $\psi_3 \rightarrow 0$
or $\pi$. This depletion is primarily a consequence of
the minimum $E_T$ requirement used to define jets.
We would expect the QCD predictions for the two angular distributions
to be very different from the phase-space model predictions.
In particular we might expect that the leading-jet
angular distribution would be similar, although not identical, to the LO
$q\overline{q} \rightarrow q\overline{q}$ scattering form.
Indeed, this is seen to be the case for both the NJETS and HERWIG
QCD predictions (Fig.~\ref{3j_three_fig}c).
We might also expect the initial-state radiation pole in the QCD matrix
element to result in an
enhanced rate of three-jet events for topologies in which the
angle between the beam direction and the three-jet plane is small.
Hence, we would expect
the $\psi_3$ distribution to be peaked towards 0 and $\pi$.
This is also evident in the HERWIG and NJETS predictions.

To prepare for the analysis of events with more than three jets we now wish
to extend the three-jet variables to describe a system of three
massive particles in the three-body rest-frame. To do this we must specify
an additional three parameters, which we take to be the single-jet mass
fractions $f_3$, $f_4$, and $f_5$ defined:
\begin{itemize}
\item[(a)] $f_{3}$, the leading-jet mass divided by the three-jet mass:
\begin{eqnarray}
f_{3} & \equiv & \frac{m_{3}}{m_{3J}}\; ,
\end{eqnarray}
\item[(b)] $f_{4}$, the next-to-leading-jet mass divided by the three-jet mass:
\begin{eqnarray}
f_{4} & \equiv & \frac{m_{4}}{m_{3J}}\; ,
\end{eqnarray}
\item[(c)] $f_{5}$, the third-to-leading-jet mass divided by the three-jet
mass:
\begin{eqnarray}
f_{5} & \equiv & \frac{m_{5}}{m_{3J}}\; .
\end{eqnarray}
\end{itemize}

HERWIG predictions for $f_3$, $f_4$, and $f_5$ are
shown in Fig.~\ref{fj_3j_fig}.
Note that $f_j$ is typically less than or of order 0.1, and
hence single-jet masses can be neglected for many purposes.

We conclude by noting that we have defined eight variables
that specify a three-jet system in the three-body rest-frame:
$m_{3J}$, $X_3$, $X_4$, $\cos\theta_3$, $\psi_3$, $f_3$, $f_4$, and $f_5$.

\end{section}
\begin{section} {Four-Jet Variables}

To completely describe a system of four jets in the four-body rest-frame
we must specify twelve independent parameters. We will choose the four-jet
mass ($m_{4J}$) and eleven dimensionless variables that span the four-body
parameter space. We have chosen a set of four-jet variables that, for
four-jet configurations that approach a three-body topology,
reduce to the three-jet variables discussed in the previous section.
This will make it possible
to compare the characteristics of four-jet events with the corresponding
characteristics of three-jet events.

The four-jet variables are shown schematically in Fig.~\ref{4jet_angle_fig}.
We begin by
reducing the four-jet system to a three-body system
by combining the two jets with the lowest two-jet mass.
We will label the two jets we combine A and B with $E_A > E_B$, where
$E_A$ and $E_B$ are the jet energies in the four-jet rest-frame.
The resulting
three-body system can be completely specified using our three-jet variables:
$X_{3'}$, $X_{4'}$, $\cos\theta_{3'}$, $\psi_{3'}$, $f_{3'}$, $f_{4'}$,
and $f_{5'}$. Note that we order the three bodies in the three-body
rest-frame so that
$E_{3'} > E_{4'} > E_{5'}$, and
use a nomenclature in which primed labels denote objects after two jets have
been combined. Hence one of the three primed objects will be the two-jet
system (AB).
Explicitly, $X_{3'}$, $X_{4'}$, $\cos\theta_{3'}$, $\psi_{3'}$, $f_{3'}$,
$f_{4'}$, and $f_{5'}$ are defined:
\begin{itemize}
\item[(i)] $X_{3'}$, the fraction of the three-body energy taken by
the leading object, normalized:
\begin{equation}
X_{3'} \; \equiv  \frac{2\;E_{3'}}{E_{3'}+E_{4'}+E_{5'}}
\equiv \frac{2\;E_{3'}}{m_{4J}} \; ,
\end{equation}
\item[(ii)] $X_{4'}$, the fraction of the three-body energy taken by
the next-to-leading object, normalized:
\begin{equation}
X_{4'} \; \equiv  \frac{2\;E_{4'}}{E_{3'}+E_{4'}+E_{5'}}
\equiv \frac{2\;E_{4'}}{m_{4J}} \; ,
\end{equation}
\item[(iii)] $\cos\theta_{3'}$, the cosine of the leading-body scattering
angle:
\begin{equation}
  \cos\theta_{3'} \;  \equiv \;  \frac{\overrightarrow{P}_{AV}\cdot
                      \overrightarrow{P}_{3'}}
                   {\mid \overrightarrow{P}_{AV}\mid
             \mid \overrightarrow{P}_{3'} \mid} \; , \\
\end{equation}
\item[(iv)] $\psi_{3'}$, the angle between the three-body plane and the
plane containing object $3'$ (the leading body) and the average beam direction:
\begin{eqnarray}
  \cos\psi_{3'}  & \equiv &
\frac{{(\overrightarrow{P}_{3'} \times
      \overrightarrow{P}_{AV})} \cdot
      {(\overrightarrow{P}_{4'} \times \overrightarrow{P}_{5'})}}
{\mid {\overrightarrow{P}_{3'} \times \overrightarrow{P}_{AV}}\mid
\mid {\overrightarrow{P}_{4'} \times \overrightarrow{P}_{5'}}\mid} \; ,
\end{eqnarray}
\item[(v)] $f_{3'}$, the mass of the leading object
divided by the four-jet mass:
\begin{eqnarray}
f_{3'} & \equiv & \frac{m_{3'}}{m_{4J}}\; ,
\end{eqnarray}
\item[(vi)] $f_{4'}$, the mass of the next-to-leading object
divided by the four-jet mass:
\begin{eqnarray}
f_{4'} & \equiv & \frac{m_{4'}}{m_{4J}}\; ,
\end{eqnarray}
\item[(vii)] $f_{5'}$, the mass of the third-to-leading object
divided by the four-jet mass:
\begin{eqnarray}
f_{5'} & \equiv & \frac{m_{5'}}{m_{4J}}\; .
\end{eqnarray}

\end{itemize}

The NJETS and HERWIG predictions for the $m_{4J}$ distribution
are shown in Fig.~\ref{m4j_fig} for four-jet events produced at the
Fermilab Proton-Antiproton Collider satisfying the requirements
$m_{4J}>650$ GeV/$c^{2}$, $\mid \cos\theta_{3'}\mid < 0.8$,
and $X_{3'} <0.9$.
The QCD predictions for the
$X_{3'}$--, $X_{4'}$--, $\cos\theta_{3'}$--, and $\psi_{3'}$--distributions are
compared with the phase-space model predictions in Fig.~\ref{4j_three_fig}.
There is reasonable agreement between the HERWIG and NJETS predictions for
all of these distributions. The QCD predictions for the
$X_{3'}$-- and $X_{4'}$--distributions are not very different from the
predictions
of the phase-space model. In contrast, the NJETS and HERWIG
$\cos\theta_{3'}$-- and $\psi_{3'}$--distributions are very different from
the more uniform phase-space model predictions.
It is interesting to compare these distributions with
the equivalent distributions for three-jet events
(Fig.~\ref{3j_three_fig}). The QCD and phase-space model predictions for
the four-jet distributions are similar but not identical to the corresponding
distributions for three-jet events.
Note that (1) in comparing the phase-space model predictions for the
$X_3$-- and $X_{3'}$--distributions we see that the predicted $X_{3'}$
distribution is depleted at large $X_{3'}$, and
(2) in comparing the phase-space model predictions for the
$X_4$-- and $X_{4'}$--distributions we see that the predicted $X_{4'}$
distribution is distorted at large $X_{4'}$.
These differences can be qualitatively understood by noting that
if $4'$ or $5'$ is the (AB)-system and hence massive
then $X_{3'} < 1$ even if  $4'$ and $5'$ are collinear.
It should also be noted that the phase-space model
$\cos\theta_{3'}$ distribution is slightly depleted at small
$\mid\cos\theta_{3'}\mid$ and the
$\psi_{3'}$ distribution is slightly depleted for values of
$\psi_{3'}$ close to 0
and $\pi$. These features are consequences of the minimum jet-jet
separation requirement
$\Delta R > 0.9$, and the minimum jet transverse energy requirement
$E_T > 20$~GeV.

The HERWIG predictions for the normalized single-jet masses $f_{j'}$ are
shown in Fig.~\ref{fj_4j_fig}. They exhibit peaks close to $f_{j'} = 0.05$
which reflect the finite single-jet masses resulting from the HERWIG
fragmentation model, and long tails at larger values of $f_{j'}$ which
reflect the contributions from the combined (AB)-systems. Note that
although single jets are massless in the NJETS calculation, the NJETS
program does predict the contribution to the $f_{j'}$--distributions
from the combined (AB)-systems, and indeed the NJETS and HERWIG
predictions are in good agreement at large $f_{j'}$.

To complete our description of the four-jet system we must now specify
four additional parameters that describe the two-jet (AB)-system.
To describe the (AB)-system we choose:
\begin{itemize}
\item[(a)] $f_{A}$, the mass of jet A divided by the four-jet mass:
\begin{eqnarray}
f_{A} & \equiv & \frac{m_{A}}{m_{4J}} \; ,
\end{eqnarray}
\item[(b)] $f_{B}$, the mass of jet B divided by the four-jet mass:
\begin{eqnarray}
f_{B} & \equiv & \frac{m_{B}}{m_{4J}} \; ,
\end{eqnarray}
\item[(c)] $X_{A}$, defined in the four-jet rest-frame as the fraction
of the energy of the (AB)-system taken by the leading jet:
\begin{eqnarray}
X_{A}  & \equiv & \frac{E_A}{E_A + E_B} \; ,
\end{eqnarray}
\item[(d)] $\psi'_{AB}$, defined in the four-jet rest-frame as the
angle between (i) the plane containing the ($AB$)-system and the average
beam direction, and (ii) the plane containing A and B
(see Fig.~\ref{4jet_angle_fig}).
The prime reminds us that in order to define $\psi'_{AB}$
we have combined two jets to obtain the (AB)-system. Note that:
\begin{eqnarray}
  \cos\psi'_{AB}  & \equiv & \frac{(\overrightarrow{P}_{A} \times
\overrightarrow{P}_{B}) \cdot
(\overrightarrow{P}_{AB} \times \overrightarrow{P}_{AV})}
{\mid \overrightarrow{P}_{A} \times \overrightarrow{P}_{B} \mid
 \mid \overrightarrow{P}_{AB} \times \overrightarrow{P}_{AV} \mid} \; .
\end{eqnarray}

\end{itemize}
The predicted $f_{A}$-- and $f_{B}$--
distributions are shown in Figs.~\ref{4j_ab_fig} (a) and
\ref{4j_ab_fig} (b) respectively.
The typical values of $f_{A}$ and $f_{B}$ predicted by the HERWIG fragmentation
model are less than or of order 0.05.
The predicted $X_{A}$
distributions are shown in Fig.~\ref{4j_ab_fig}~(c).
The NJETS and HERWIG QCD calculations yield harder
$X_{A}$ distributions than the corresponding distribution predicted
by the phase-space model. Presumably this reflects the presence of the soft
gluon radiation pole in the QCD matrix element.
To gain some insight into the shape of the phase-space model prediction
for the $X_{A}$ distribution consider a system of four massless particles
labelled randomly i, j, k, and l. If we define $X_i \equiv E_i / (E_i + E_j)$,
then the phase-space prediction for the distribution of events as a function
of $X_i$ is given by:
\begin{eqnarray}
\frac{dN}{d X_i} & \sim & \frac{3}{X_i^2} - \frac{1}{X_i^3} - 2 \; .
\end{eqnarray}
This function is already quite similar to the phase-space model prediction
shown in Fig.~\ref{4j_ab_fig}~(c), which is obtained by requiring that
the (AB)-system is the lowest mass pair, and taking account of finite
single-jet masses and experimental selection requirements.
Finally, the predicted $\psi'_{AB}$ distributions are shown in
Fig.~\ref{4j_ab_fig}~(d).
The NJETS and HERWIG predictions for the $\psi'_{AB}$
distribution are in agreement with one another.
The slight decrease in the population of events predicted by the
phase-space model as $\psi'_{AB}$ approaches 0 or $\pi$
is a consequence of the minimum jet $E_T$ requirement.

We conclude by noting that we have defined twelve variables
that specify a four-jet system in the four-body rest-frame:
$m_{4J}$, $X_{3'}$, $X_{4'}$, $\cos\theta_{3'}$, $\psi_{3'}$,
$f_{3'}$, $f_{4'}$, $f_{5'}$, $f_{A}$, $f_{B}$, $X_{A}$,
and $\psi'_{AB}$.

\end{section}
\begin{section} {Five-Jet Variables}

To completely describe a system of five jets in the five-body rest-frame
we must specify sixteen independent parameters. We will choose the five-jet
mass ($m_{5J}$) and fifteen dimensionless variables that span the five-body
parameter space. We have chosen a set of five-jet variables that, for
five-body configurations that approach a four-body topology,
reduce to the four-jet variables discussed in the previous section.
Furthermore, for
five-body configurations that approach a three-body topology, our
five-jet parameters
reduce to the three-jet variables discussed previously.
Thus we will be able
to compare the characteristics of five-jet events with the corresponding
characteristics of three-jet and four-jet events.

The five-jet variables are shown schematically in Fig.~\ref{5jet_angle_fig}.
We begin by
reducing the five-jet system to a four-body system
by combining the two jets with the lowest two-jet mass.
We will label the two jets we combine C and D, with $E_C > E_D$, where
$E_C$ and $E_D$ are the jet energies in the five-jet rest-frame.
We can then further reduce the resulting four-body system to a three-body
system by combining the two bodies with the lowest two-body mass.
We will label the two objects we combine $A'$ and $B'$, with $E_{A'} > E_{B'}$.
The resulting
three-body system can be completely specified using our three-jet variables:
$X_{3''}$, $X_{4''}$, $\cos\theta_{3''}$, $\psi_{3''}$, $f_{3''}$, $f_{4''}$,
and $f_{5''}$. Note that we order the three bodies such that
$E_{3''} > E_{4''} > E_{5''}$, and
use a nomenclature in which doubly primed labels denote objects after
two operations in which the two bodies with the lowest two-body
mass have been combined.
One of the three doubly primed objects will be the ($A'B'$)-system.
Explicitly, $X_{3''}$, $X_{4''}$, $\cos\theta_{3''}$, $\psi_{3''}$, $f_{3''}$,
$f_{4''}$, and $f_{5''}$ are defined:
\begin{itemize}
\item[(i)] $X_{3''}$, the fraction of the three-body energy taken by
the leading body, normalized:
\begin{equation}
X_{3''}  \equiv \; \frac{2\;E_{3''}}{E_{3''}+E_{4''}+E_{5''}}
= \; \frac{2E_{3''}}{m_{5J}}\; ,
\end{equation}
\item[(ii)] $X_{4''}$, the fraction of the three-body energy taken by
the next-to-leading body, normalized:
\begin{equation}
X_{4''}  \equiv \; \frac{2\;E_{4''}}{E_{3''}+E_{4''}+E_{5''}}
= \; \frac{2E_{4''}}{m_{5J}} \; ,
\end{equation}
\item[(iii)] $\cos\theta_{3''}$, the cosine of the leading-body
scattering angle:
\begin{equation}
  \cos\theta_{3''} \; \equiv \; \frac{\overrightarrow{P}_{AV}\cdot
                      \overrightarrow{P}_{3''}}
                   {\mid \overrightarrow{P}_{AV}\mid
             \mid \overrightarrow{P}_{3''} \mid} \; , \\
\end{equation}
\item[(iv)] $\psi_{3''}$, the angle between the three-body plane and the
plane containing object $3''$ (the leading body) and the average beam
direction:
\begin{eqnarray}
  \cos\psi_{3''}  & \equiv &
\frac{{(\overrightarrow{P}_{3''} \times
      \overrightarrow{P}_{AV})} \cdot
      {(\overrightarrow{P}_{4''} \times \overrightarrow{P}_{5''})}}
{\mid {\overrightarrow{P}_{3''} \times \overrightarrow{P}_{AV}}\mid
\mid {\overrightarrow{P}_{4''} \times \overrightarrow{P}_{5''}}\mid} \; ,
\end{eqnarray}
\item[(v)] $f_{3''}$, the normalized mass of the leading object:
\begin{eqnarray}
f_{3''} & \equiv & \frac{m_{3''}}{m_{5J}}\; ,
\end{eqnarray}
\item[(vi)] $f_{4''}$, the normalized mass of the next-to-leading object:
\begin{eqnarray}
f_{4''} & \equiv & \frac{m_{4''}}{m_{5J}}\; ,
\end{eqnarray}
\item[(vii)] $f_{5''}$, the normalized mass of the third-to-leading object:
\begin{eqnarray}
f_{5''} & \equiv & \frac{m_{5''}}{m_{5J}}\; .
\end{eqnarray}

\end{itemize}

The NJETS and HERWIG predictions for the $m_{5J}$ distribution
are shown in Fig.~\ref{m5j_fig} for five-jet events produced at the
Fermilab Proton-Antiproton Collider and satisfying the requirement
$m_{5J} > 750$~GeV/c$^2$. The QCD predictions for the
$X_{3''}$--, $X_{4''}$--, $\cos\theta_{3''}$--, and $\psi_{3''}$--distributions
are
compared with the phase-space model predictions in Fig.~\ref{5j_three_fig}.
The predicted distributions are qualitatively similar to the equivalent
four-jet distributions shown in Fig.~\ref{4j_three_fig}.
Note that the QCD predictions for
the $\cos\theta_{3''}$ distribution are remarkably similar to the
simple LO $q\overline{q} \rightarrow q\overline{q}$ angular distribution.

The HERWIG predictions for the normalized single-jet masses $f_{j''}$ are
shown in Fig.~\ref{fj_5j_fig}. Once again, the HERWIG and NJETS distributions
are in agreement at large mass fractions.

We must now specify the intermediate four-body system. In analogy with the
four-jet analysis we will do this by specifying four additional
dimensionless variables that describe the ($A'B'$)-system. We choose $f_{A'}$,
$f_{B'}$, $X_{A'}$, and $\psi''_{A'B'}$, defined:
\begin{itemize}
\item[(a)] $f_{A'}$, the normalized mass of object $A'$:
\begin{eqnarray}
f_{A'} & \equiv & \frac{m_{A'}}{m_{5J}} \; ,
\end{eqnarray}
\item[(b)] $f_{B'}$, the normalized mass of object $B'$:
\begin{eqnarray}
f_{B'} & \equiv & \frac{m_{B'}}{m_{5J}} \; ,
\end{eqnarray}
\item[(c)] $X_{A'}$, defined in the five-jet rest-frame as the fraction
of the energy of the ($A'B'$)-system taken by the leading body:
\begin{eqnarray}
X_{A'}  & \equiv & \frac{E_{A'}}{E_{A'} + E_{B'}} \; ,
\end{eqnarray}
\item[(d)] $\psi''_{A'B'}$, defined in the five-jet rest-frame as the
angle between (i) the plane containing the ($A'B'$)-system and the average beam
direction, and (ii) the plane containing $A'$ and $B'$
(see Fig.~\ref{5jet_angle_fig}).
Note that:
\begin{eqnarray}
  \cos\psi''_{A'B'}  & \equiv & \frac{(\overrightarrow{P}_{A'} \times
\overrightarrow{P}_{B'}) \cdot
(\overrightarrow{P}_{A'B'} \times \overrightarrow{P}_{AV})}
{\mid \overrightarrow{P}_{A'} \times \overrightarrow{P}_{B'} \mid
 \mid \overrightarrow{P}_{A'B'} \times \overrightarrow{P}_{AV} \mid} \; .
\end{eqnarray}

\end{itemize}
The predicted distributions of these variables are shown in
Fig.~\ref{5j_ab_fig}.
The HERWIG predictions for the
$f_{A'}$-- and $f_{B'}$--distributions peak at values of about 0.02 and
have long tails associated with composite $A'$ or $B'$ systems. The tails
are accounted for by the NJETS predictions. It is interesting to compare
the $X_{A'}$-- and $\psi''_{A'B'}$--distributions with the
corresponding four-jet distributions (Figs.~\ref{4j_ab_fig} (c) and
\ref{4j_ab_fig} (d) respectively). The QCD and phase-space model predictions
for the
five-jet distributions are qualitatively similar to the corresponding
four-jet distributions. Note that the HERWIG and NJETS
predictions are in agreement with one another.

Finally, to complete our specification of the five-jet system we must
define a further four variables that describe the two-body (CD)-system.
We choose $f_{C}$, $f_{D}$, $X_{C}$, and $\psi''_{CD}$,
defined:
\begin{itemize}
\item[(a)] $f_{C}$, the normalized mass of jet C:
\begin{eqnarray}
f_{C} & \equiv & \frac{m_{C}}{m_{5J}} \; ,
\end{eqnarray}
\item[(b)] $f_{D}$, the normalized mass of jet D:
\begin{eqnarray}
f_{D} & \equiv & \frac{m_{D}}{m_{5J}} \; ,
\end{eqnarray}
\item[(c)] $X_{C}$, defined in the five-jet rest-frame as the fraction
of the energy of the (CD)-system taken by the leading jet:
\begin{eqnarray}
X_{C}  & \equiv & \frac{E_C}{E_C + E_D} \; ,
\end{eqnarray}
\item[(d)] $\psi''_{CD}$, defined in the five-jet rest-frame as the
angle between (i) the plane containing the ($CD$)-system and the average beam
direction and (ii) the plane containing $C$ and $D$
(see Fig.~\ref{5jet_angle_fig}).
Note that:
\begin{eqnarray}
  \cos\psi''_{CD}  & \equiv & \frac{(\overrightarrow{P}_{C} \times
\overrightarrow{P}_{D}) \cdot
(\overrightarrow{P}_{CD} \times \overrightarrow{P}_{AV})}
{\mid \overrightarrow{P}_{C} \times \overrightarrow{P}_{D} \mid
 \mid \overrightarrow{P}_{CD} \times \overrightarrow{P}_{AV} \mid} \; .
\end{eqnarray}

\end{itemize}
The predicted distributions of these variables are shown in
Fig.~\ref{5j_cd_fig}.
The HERWIG predictions for the
$f_{A'}$-- and $f_{B'}$--distributions peak at values less than 0.02.
Note that the QCD predictions for the $X_{C}$ distribution are harder
than the corresponding phase-space model prediction,
whilst the QCD predictions for the
$\psi''_{CD}$--distribution are similar to the corresponding
phase-space model prediction.

We conclude by noting that we have defined sixteen variables
that specify a five-jet system in the five-body rest-frame:
$m_{5J}$, $X_{3''}$, $X_{4''}$, $\cos\theta_{3''}$, $\psi_{3''}$,
$f_{3''}$, $f_{4''}$, $f_{5''}$, $f_{A'}$, $f_{B'}$, $X_{A'}$,
$\psi''_{A'B'}$, $f_{C}$, $f_{D}$, $X_{C}$, and
$\psi''_{CD}$.

\end{section}
\begin{section} {Generalization to Events with Six or More Jets}

A list of the multijet variables described in the preceeding sections
is given in Table~\ref{variables_tab}. The extension of the variables
to describe multijet systems with more than five jets is straight forward.
As an example the variables required to describe a six-jet event are also
listed in Table~\ref{variables_tab}.
In general, to describe an event containing N jets
we use the mass of the N-jet system plus (4N-5) dimensionless variables.
To define the dimensionless variables we proceed by reducing the N-jet
system to a three-body system. This is done in (N-3) steps.
In each step the two bodies with the lowest two-body mass are combined by
adding the two four-vectors. The resulting
three-body system is described by specifying seven parameters, namely the
normalized masses of the three bodies (e.g. $f_3$, $f_4$, and $f_5$),
the Dalitz variables for the two leading bodies (e.g. $X_3$ and $X_4$),
the cosine of the leading-body scattering angle (e.g. $\cos\theta_3$),
and the angle between the three-body plane and the beam direction
(e.g. $\psi_3$). To complete the description of the N-jet system we must
then specify an additional four parameters for each step in which two
bodies were combined. These parameters are the normalized masses of the
two bodies (e.g. $f_A$ and $f_B$), the fraction of the two-body energy
taken by the leading body (e.g. $X_A$), and the
angle defined in the N-jet rest-frame between the plane containing the
two-body system and the beam direction and the plane defined by the two
bodies (e.g $\psi'_{AB}$).

\end{section}
\begin{section} {Summary}

We have defined a set of (4N~-~4) parameters that can be used to analyse
events in which N jets have been produced in high energy hadron-hadron
collisions. These multijet parameters
(i) span the multijet parameter space,
(ii) facilitate the interpretation of observed event distributions within
the framework of perturbative QCD, and
(iii) make it is possible to compare the characteristics of events having
N jets with the characteristics of events having for example (N+1) jets.

To illustrate the use of the multijet variables described in this paper
we have discussed QCD and phase-space model predictions for three-jet,
four-jet, and five-jet events produced at the Fermilab Proton-Antiproton
Collider. For this particular example we note that,
apart from small effects that can be ascribed to the absence of a
fragmentation model in the NJETS calculation,
the complete LO QCD matrix element predictions for each of the multijet
parameter
distributions discussed in the preceeding sections are well described by the
parton shower Monte Carlo calculation. Thus it appears that even when there
are five hard partons in the final state a good approximation to the
LO QCD matrix
element is given by $2 \rightarrow 2$ scattering plus gluon radiation.
This is of interest because the complete LO matrix element calculation is
not at present available for topologies with more than five final state
partons.
Hence for the analysis of events with six or more jets we must rely on
parton shower Monte Carlo calculations, or on other approximations to the
QCD matrix element.
%

Finally, the multijet variables discussed in this paper have been
selected to emphasize simple to interpret quantities (masses,
energy fractions, scattering angles, and planarity-type angles).
Wherever possible we have tried to select parameters for which the
phase-space model distributions are simple to understand. Experimental
requirements used to select well measured N-jet events necessarily distort
some of the predicted distributions.
However, for the example discussed in this paper, we note that
for most parameters the experimental selection criteria result in
modifications to the phase-space model predictions that are modest
and are limited to small regions of the parameter space. The observed
N-jet distributions should therefore directly reflect the dynamics of
the underlying multijet matrix element.

\end{section}

\vspace{0.3cm}

We are grateful to Walter Giele for many interesting discussions.
This work was supported by the U.S. Department of Energy,
and the Ministry of Science, Culture and Education of Japan.

\vspace{0.2cm}
%

\clearpage
\begin{table}
\begin{center}
\begin{tabular}{|c|c|c|c|c|}
\hline
\hline
Two-Jet&Three-Jet&Four-Jet&Five-Jet&Six-Jet\\
\hline
$m_{2J}$ & $m_{3J}$ & $m_{4J}$ & $m_{5J}$ & $m_{6J}$ \\
$\cos\theta^{\star}$&$\cos\theta_3$&$\cos\theta_{3'}$&$\cos\theta_{3''}$
&$\cos\theta_{3'''}$\\
$f_3$ & $f_3$ & $f_{3'}$ & $f_{3''}$ & $f_{3'''}$ \\
$f_4$ & $f_4$ & $f_{4'}$ & $f_{4''}$ & $f_{4'''}$ \\
\hline
      & $f_5$ & $f_{5'}$ & $f_{5''}$ & $f_{5'''}$ \\
      &$\psi_3$&$\psi_{3'}$&$\psi_{3''}$&$\psi_{3'''}$\\
      &$X_3$&$X_{3'}$&$X_{3''}$&$X_{3'''}$\\
      &$X_4$&$X_{4'}$&$X_{4''}$&$X_{4'''}$\\
\hline
      &     &$f_{A}$&$f_{A'}$&$f_{A''}$\\
      &     &$f_{B}$&$f_{B'}$&$f_{B''}$\\
      &     &$X_{A}$&$X_{A'}$&$X_{A''}$\\
      &     &$\psi'_{AB}$&$\psi''_{A'B'}$&$\psi'''_{A''B''}$\\
\hline
      &     &    &$f_{C}$&$f_{C'}$\\
      &     &    &$f_{D}$&$f_{D'}$\\
      &     &    &$X_{C}$&$X_{C'}$\\
      &     &    &$\psi''_{CD}$&$\psi'''_{C'D'}$\\
\hline
      &     &    &   &$f_{E}$\\
      &     &    &   &$f_{F}$\\
      &     &    &   &$X_{E}$\\
      &     &    &   &$\psi'''_{EF}$\\
\hline
\hline
\end{tabular}
\vspace{0.5cm}
\caption{Summary of the (4N-4) multijet variables for N = 2, 3, 4, 5, and 6.}
\label{variables_tab}
\end{center}
\end{table}
%
\clearpage
%
\clearpage
\clearpage
\vspace{-2.5in}
\clearpage
\begin{figure}
\vspace{0.5cm}
\epsfysize=3.0in
\epsffile[100 144 522 590]{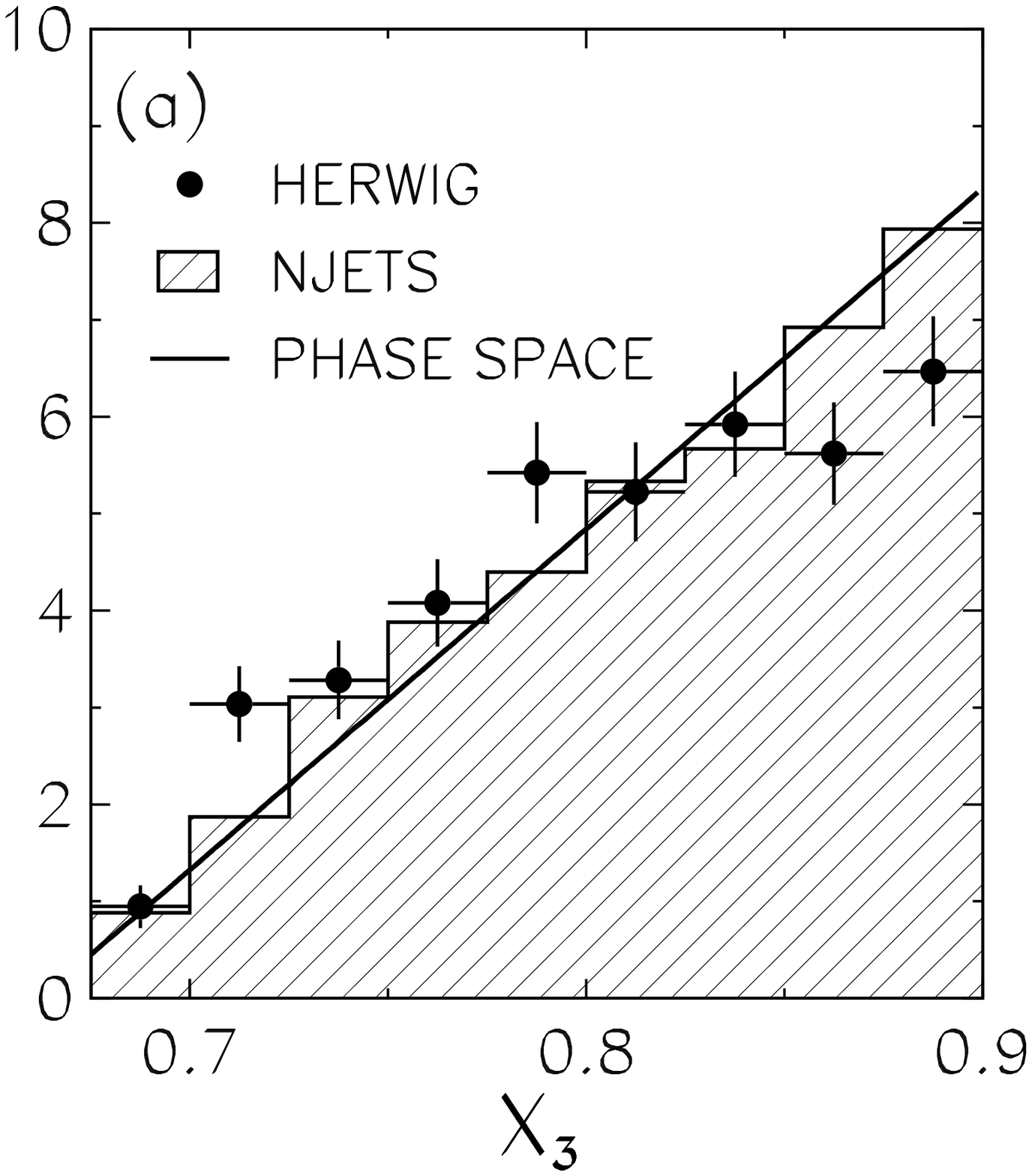}
\vspace{-3.0in}
\epsfysize=3.0in
\epsffile[-410 144 522 590]{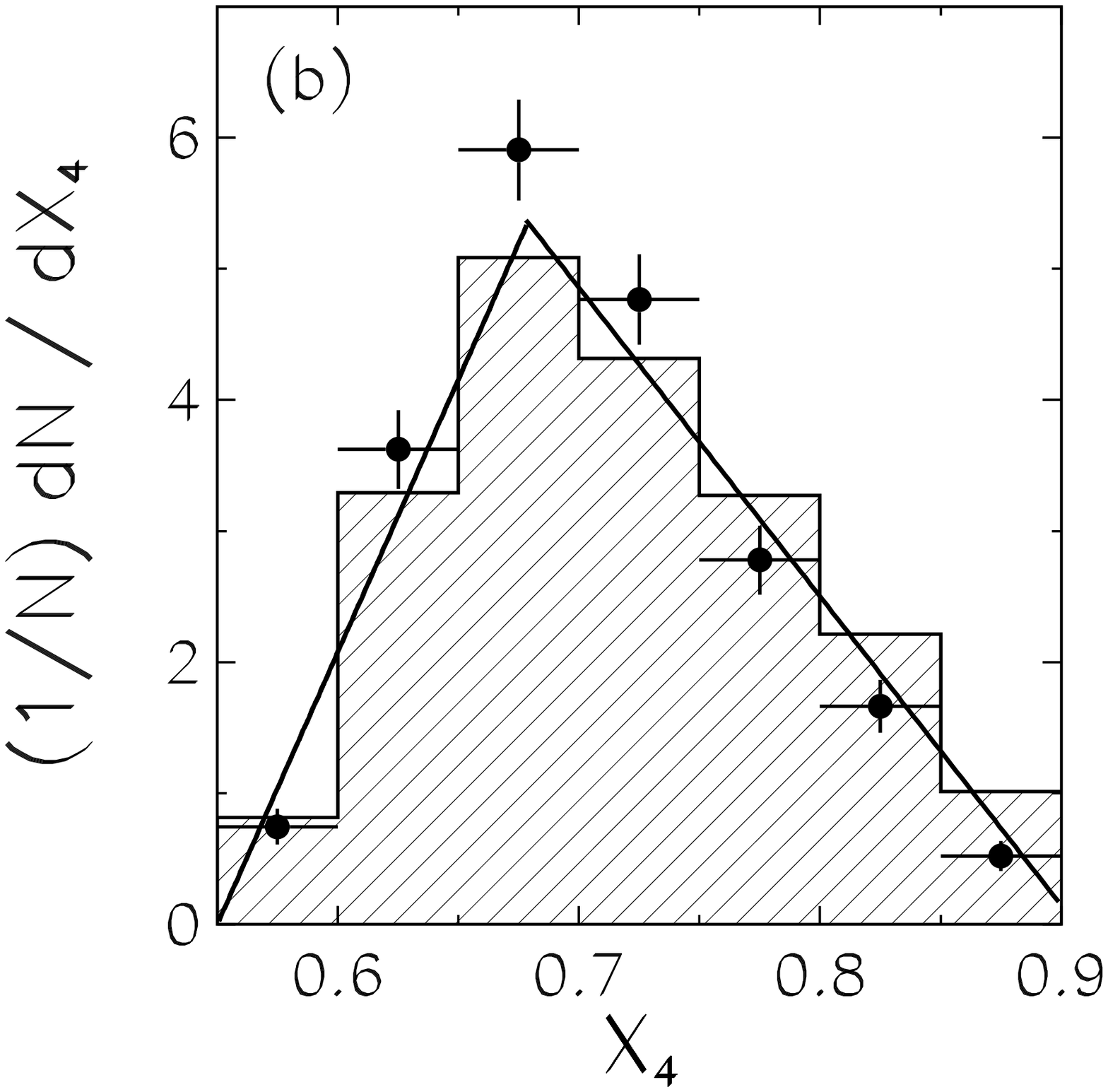}
\end{figure}
\vspace{1cm}
\begin{figure}
\epsfysize=3.0in
\epsffile[100 144 522 590]{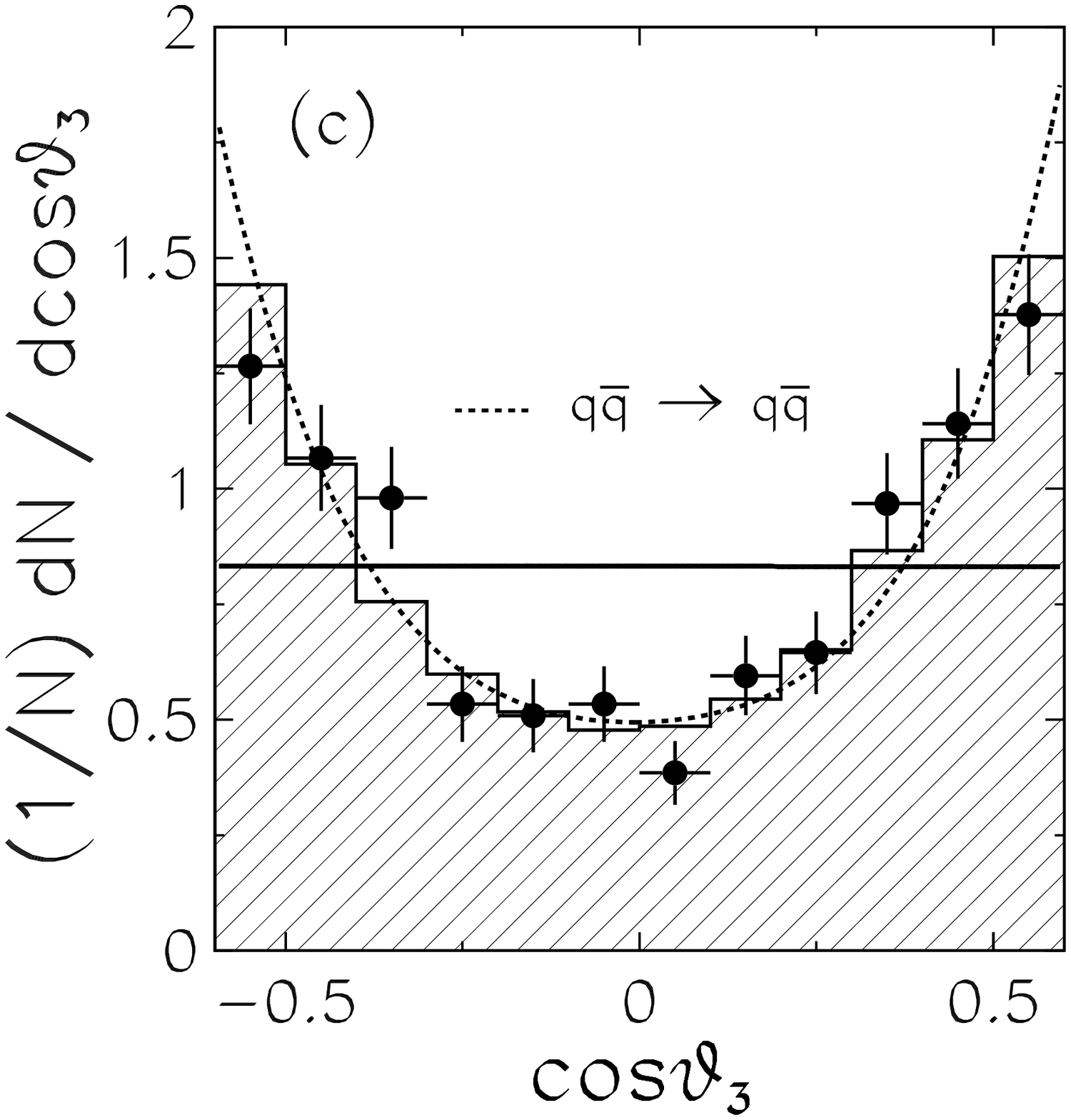}
\vspace{-3.0in}
\epsfysize=3.0in
\epsffile[-410 144 522 590]{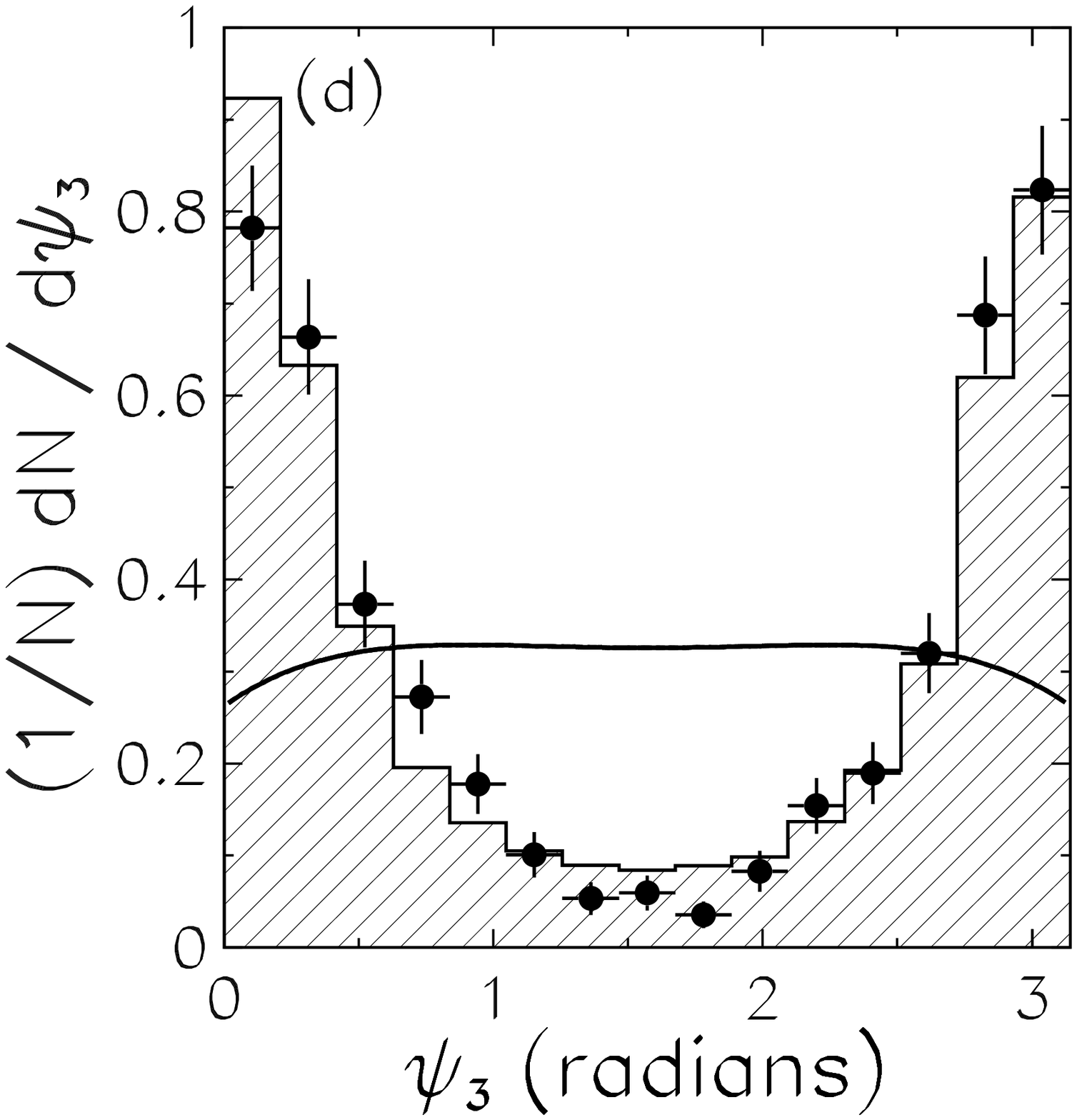}
\caption{Predicted distributions of the three-jet variables defined in the text
for three-jet events produced at the Fermilab Proton-Antiproton Collider
that satisfy the
requirements $m_{3J}>600$~GeV/$c^{2}$, $X_3<0.9$, and
$\mid \cos\theta_{3}\mid < 0.6$.
HERWIG predictions (points) are compared with NJETS predictions (histograms)
and the phase-space model predictions (solid curves) for
(a) $X_3$, (b) $X_4$, (c) $\cos \theta_{3}$, and (d) $\psi_3$.
The broken curve in the $\cos \theta_{3}$ figure is the LO QCD prediction
for $q\overline{q} \rightarrow q\overline{q}$ scattering.}
\label{3j_three_fig}
\end{figure}
\begin{figure}
\epsfysize=5.0in
\vspace{0.5cm}
\epsffile[30 144 522 590]{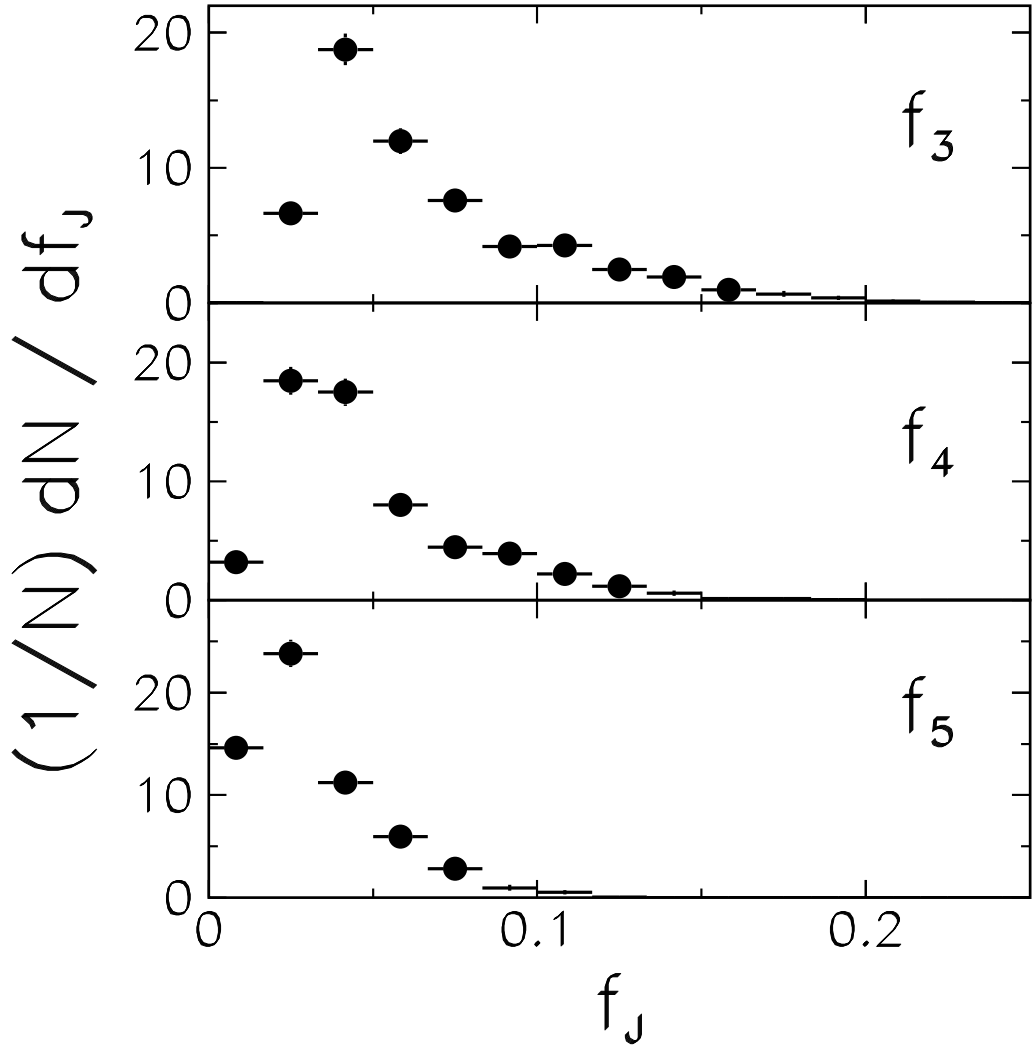}
\caption{HERWIG Monte Carlo predictions for the
single-jet mass-fraction distributions for jets in three-jet events
produced at the Fermilab Proton-Antiproton Collider that
satisfy the requirements $m_{3J}>600$ GeV/$c^{2}$,
$X_3 < 0.9$, and $\mid \cos\theta_3\mid < 0.6$.}
  \label{fj_3j_fig}
\end{figure}
\clearpage
\begin{figure}
 \epsfysize=6.0in
 \vspace{-3in}
 \epsffile[-50 144 522 848]{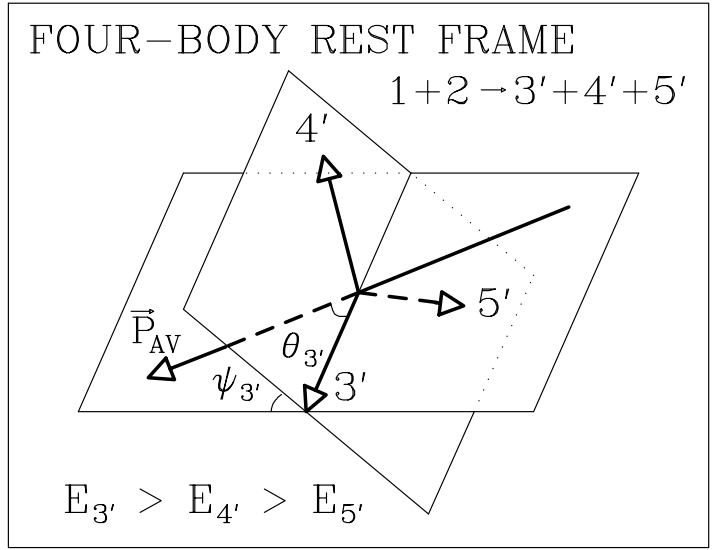}
 \epsfysize=6.0in
 \vspace{-2.5in}
 \epsffile[ -50 144 522 848]{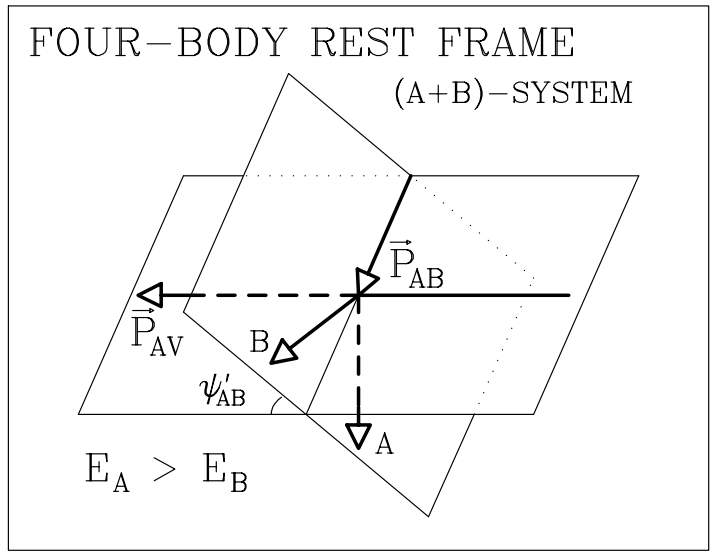}
\vspace{-1.0cm}
 \caption{Schematic definition of angles used to describe the four-jet
system in the four-jet rest-frame.}
 \label{4jet_angle_fig}
\end{figure}
\clearpage
\begin{figure}
\epsfysize=3.0in
\vspace{0.5cm}
\epsffile[-120 144 522 590]{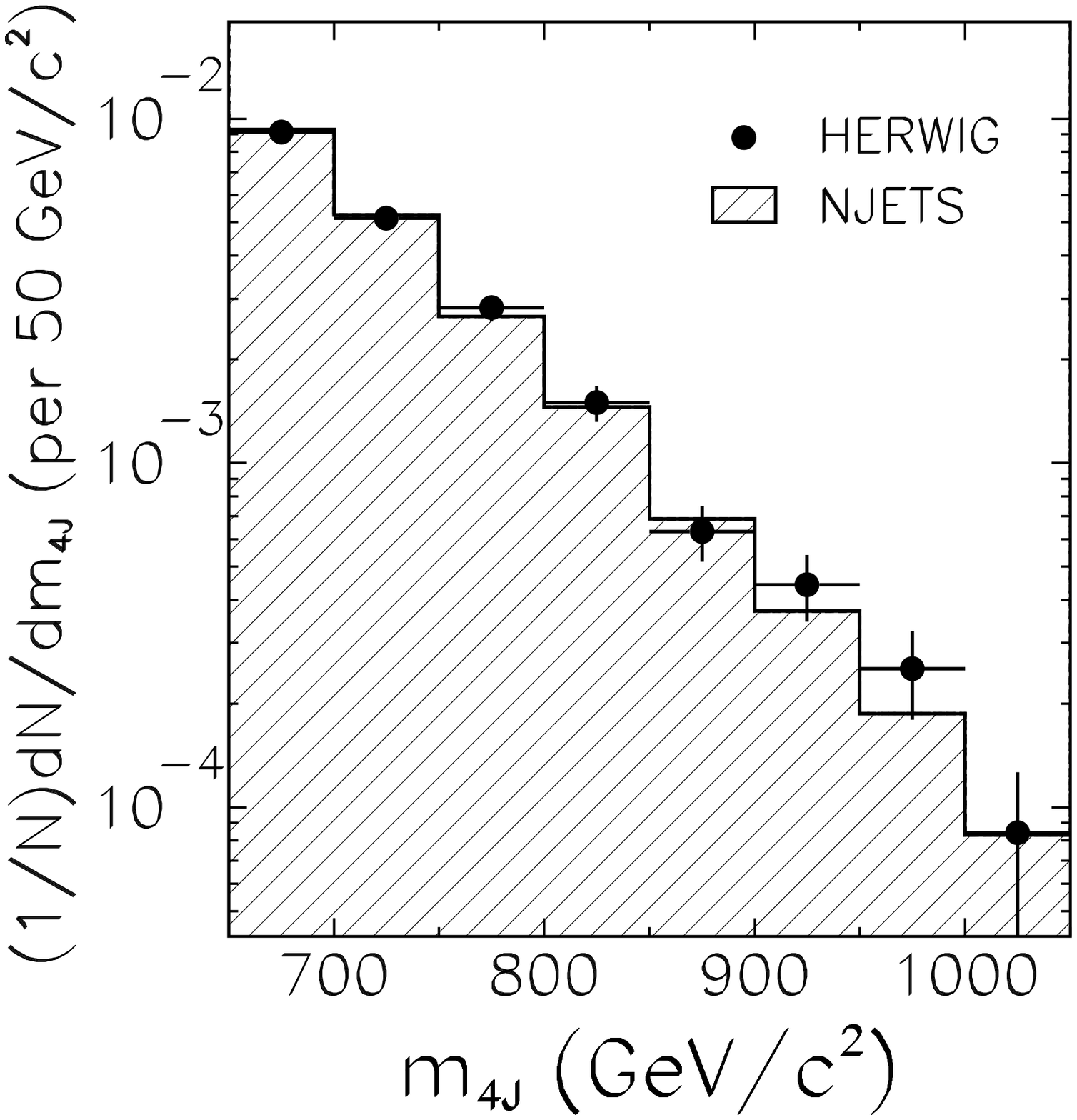}
\caption{Predicted four-jet mass distributions for events
produced at the Fermilab Proton-Antiproton Collider that satisfy the
requirements $m_{4J}>650$~GeV/$c^{2}$, $X_{3'}<0.9$, and
$\mid\cos\theta_{3'}\mid<0.8$. HERWIG predictions (points)
are compared with
NJETS predictions (histogram).}
  \label{m4j_fig}
\end{figure}
\clearpage
\begin{figure}
\vspace{0.5cm}
\epsfysize=3.0in
\epsffile[100 144 522 590]{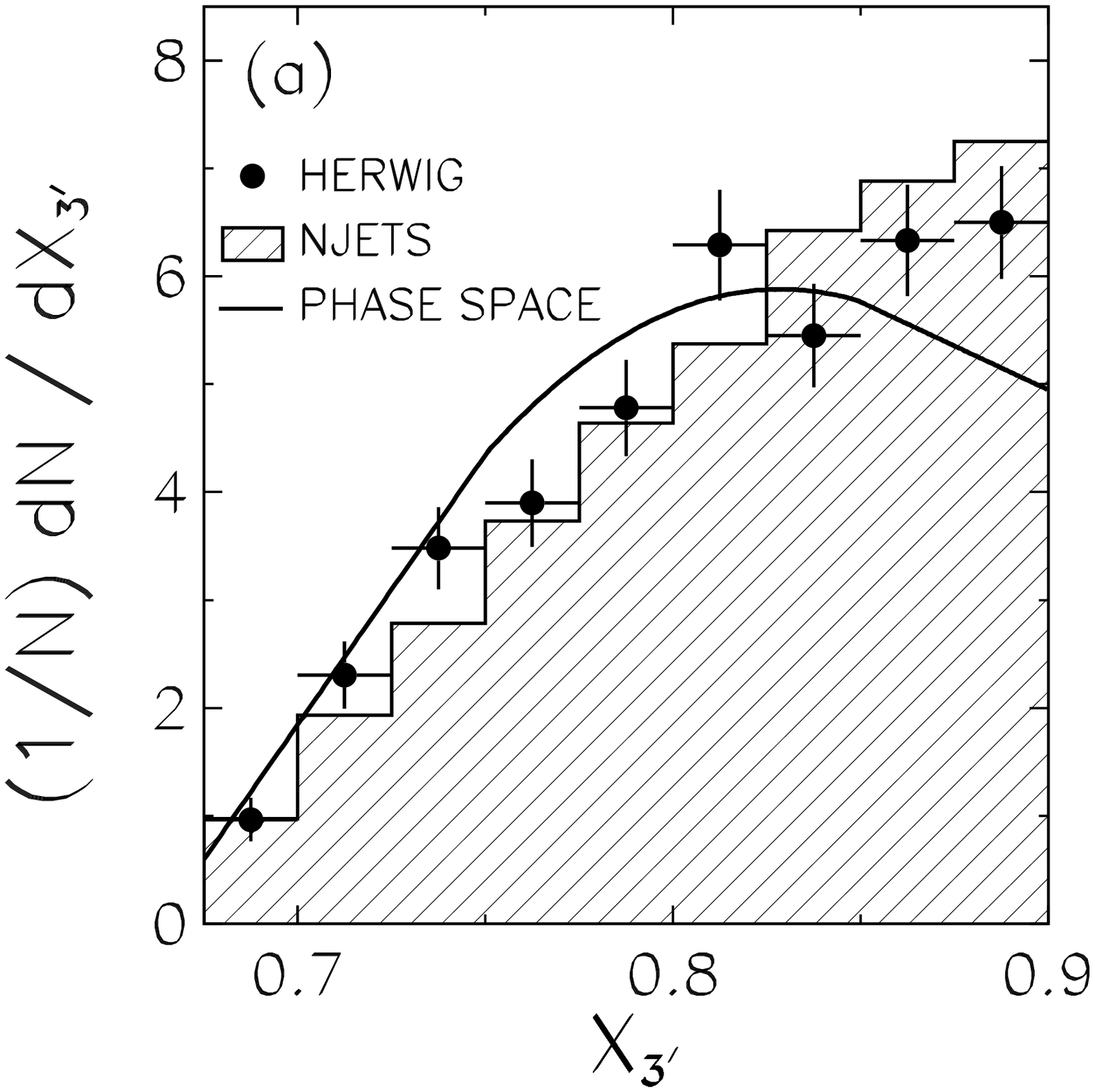}
\vspace{-3.0in}
\epsfysize=3.0in
\epsffile[-410 144 522 590]{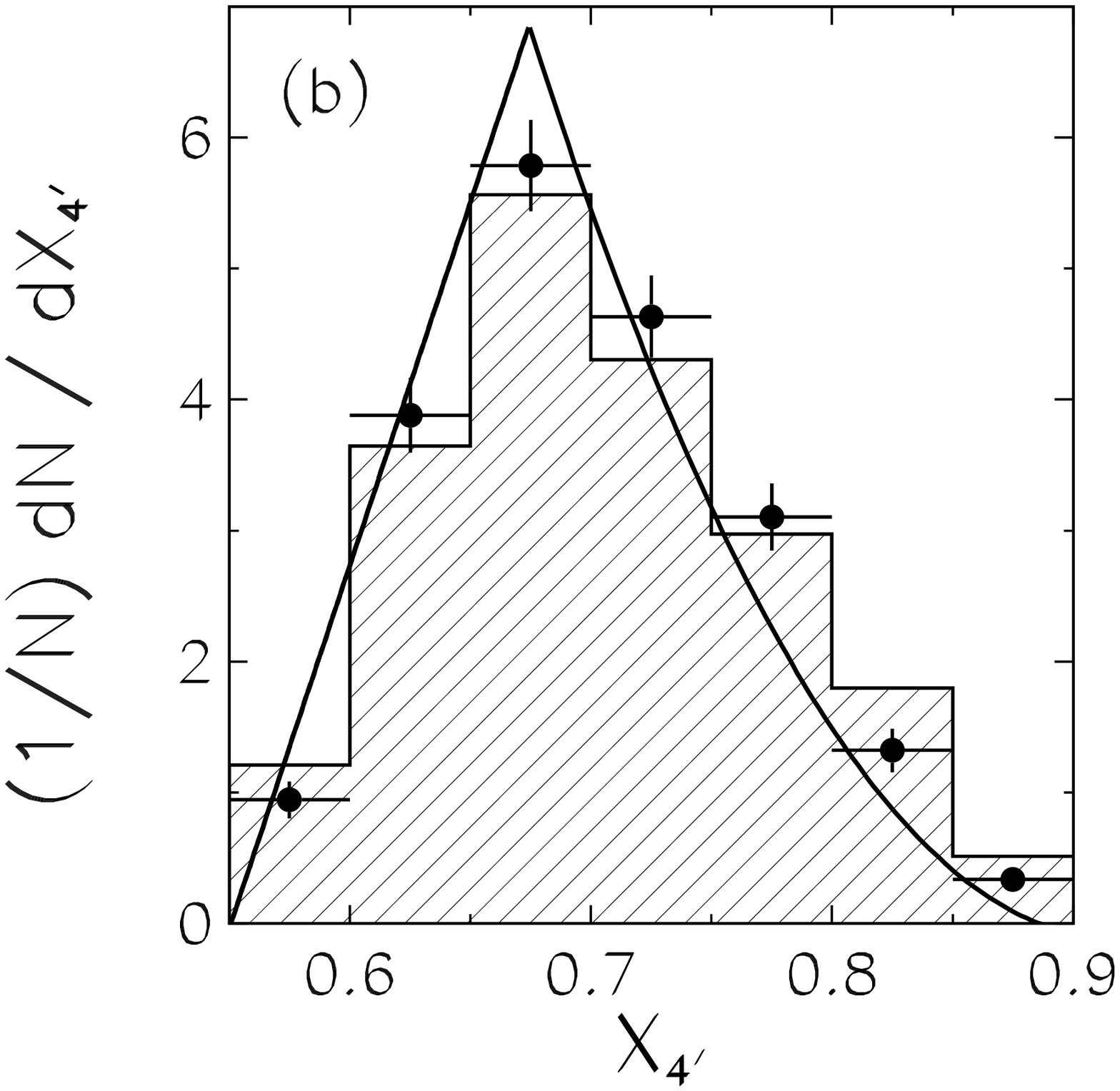}
\end{figure}
\vspace{1cm}
\begin{figure}
\epsfysize=3.0in
\epsffile[100 144 522 590]{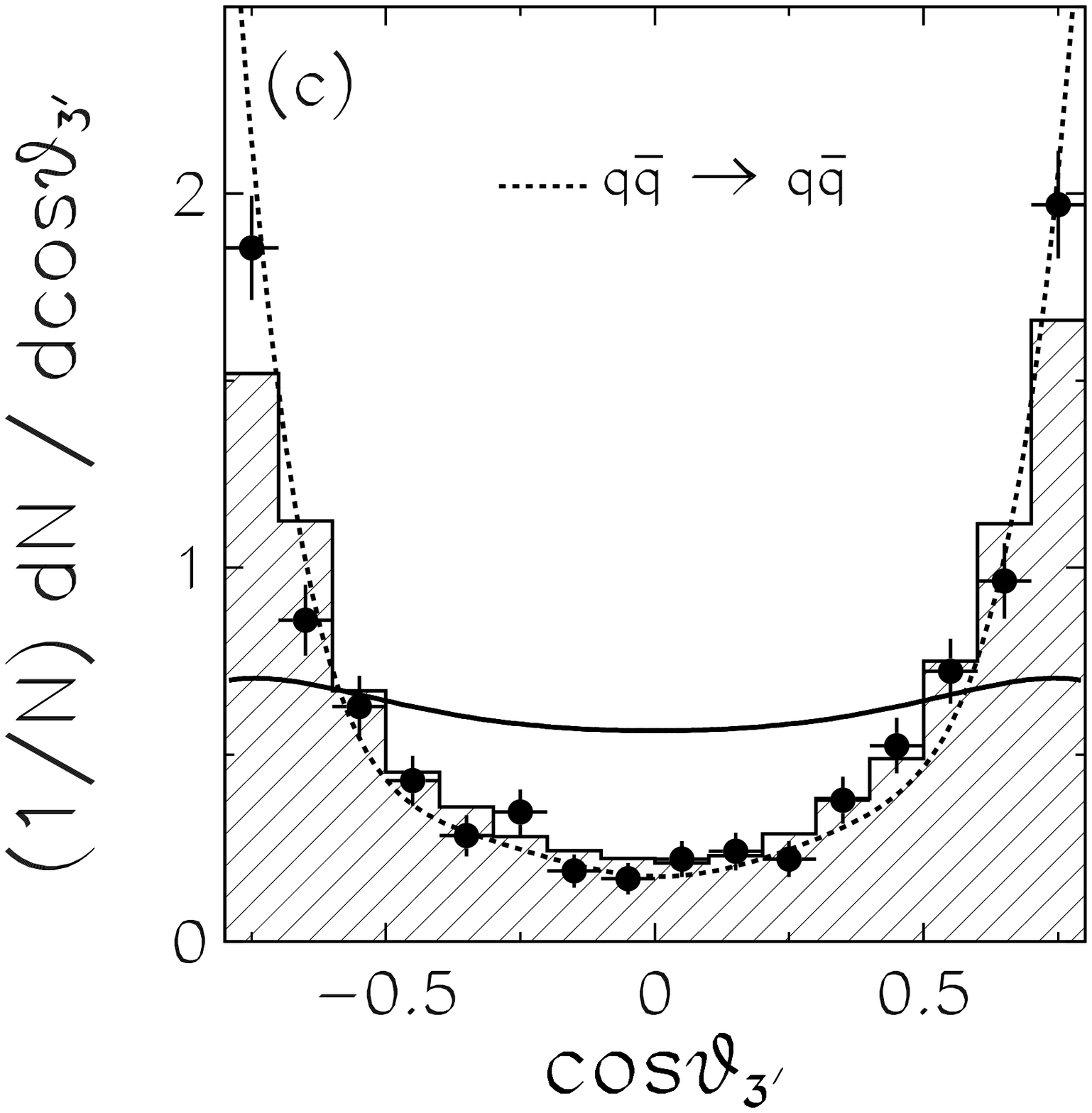}
\vspace{-3.0in}
\epsfysize=3.0in
\epsffile[-410 144 522 590]{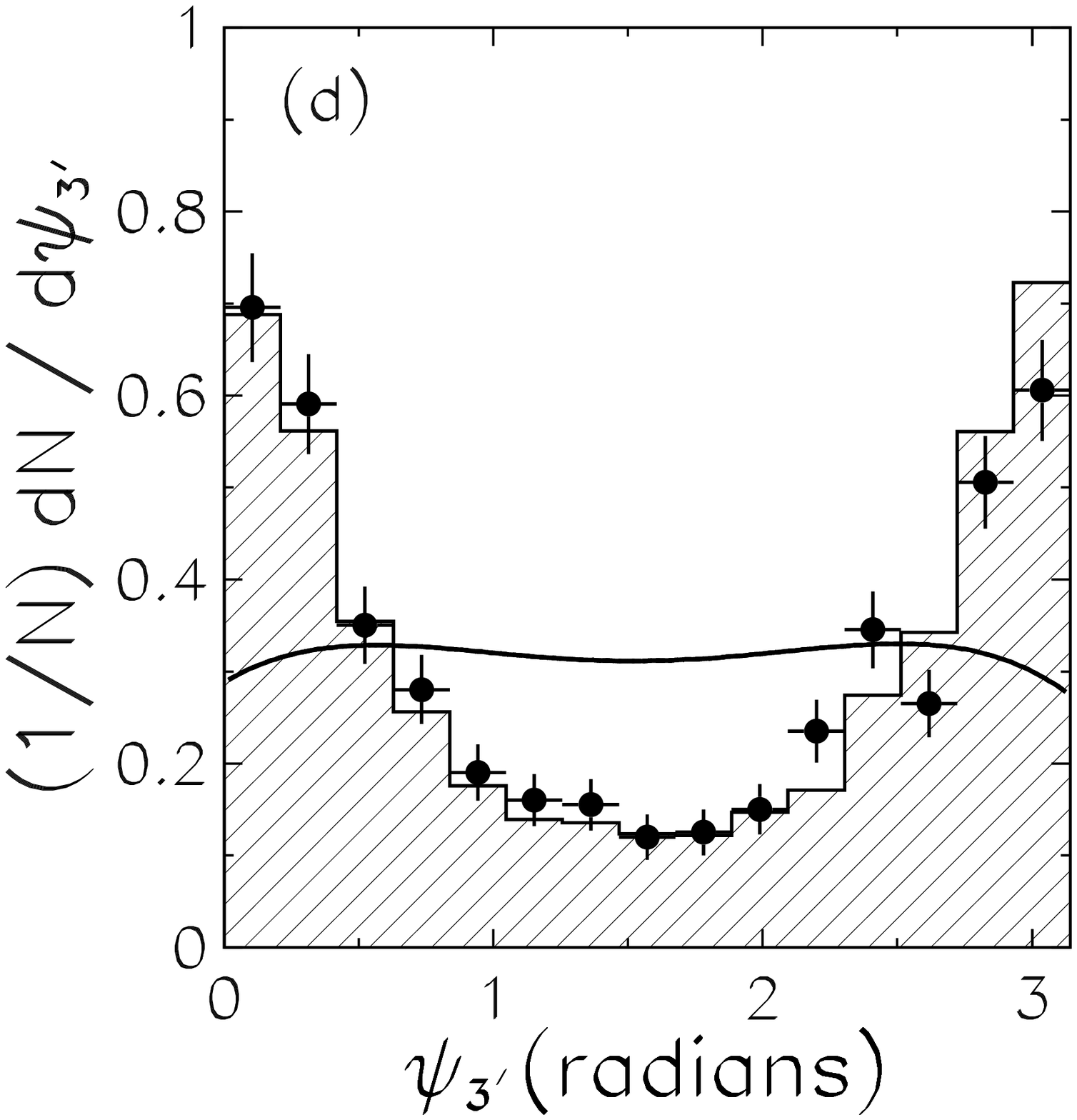}
\caption{Predicted distributions of three-body variables described in the
text for four-jet events
produced at the Fermilab Proton-Antiproton Collider that satisfy
the requirements
$m_{4J}> 650$ GeV/$c^{2}$, $X_{3'} < 0.9$, and
$\mid \cos\theta_{3'}\mid < 0.8$.
The HERWIG predictions (points) are compared with NJETS predictions
(histograms), and with the phase-space model predictions (solid curves) for
(a) $X_{3'}$, (b) $X_{4'}$, (c)
$\cos \theta_{3'}$, and (d) $\psi_{3'}$.
The broken curve in the $\cos \theta_{3'}$ figure is the LO QCD prediction
for $q\overline{q} \rightarrow q\overline{q}$ scattering.}
\label{4j_three_fig}
\end{figure}
\clearpage
\begin{figure}
\epsfysize=5.0in
\vspace{0.5cm}
\epsffile[30 144 522 590]{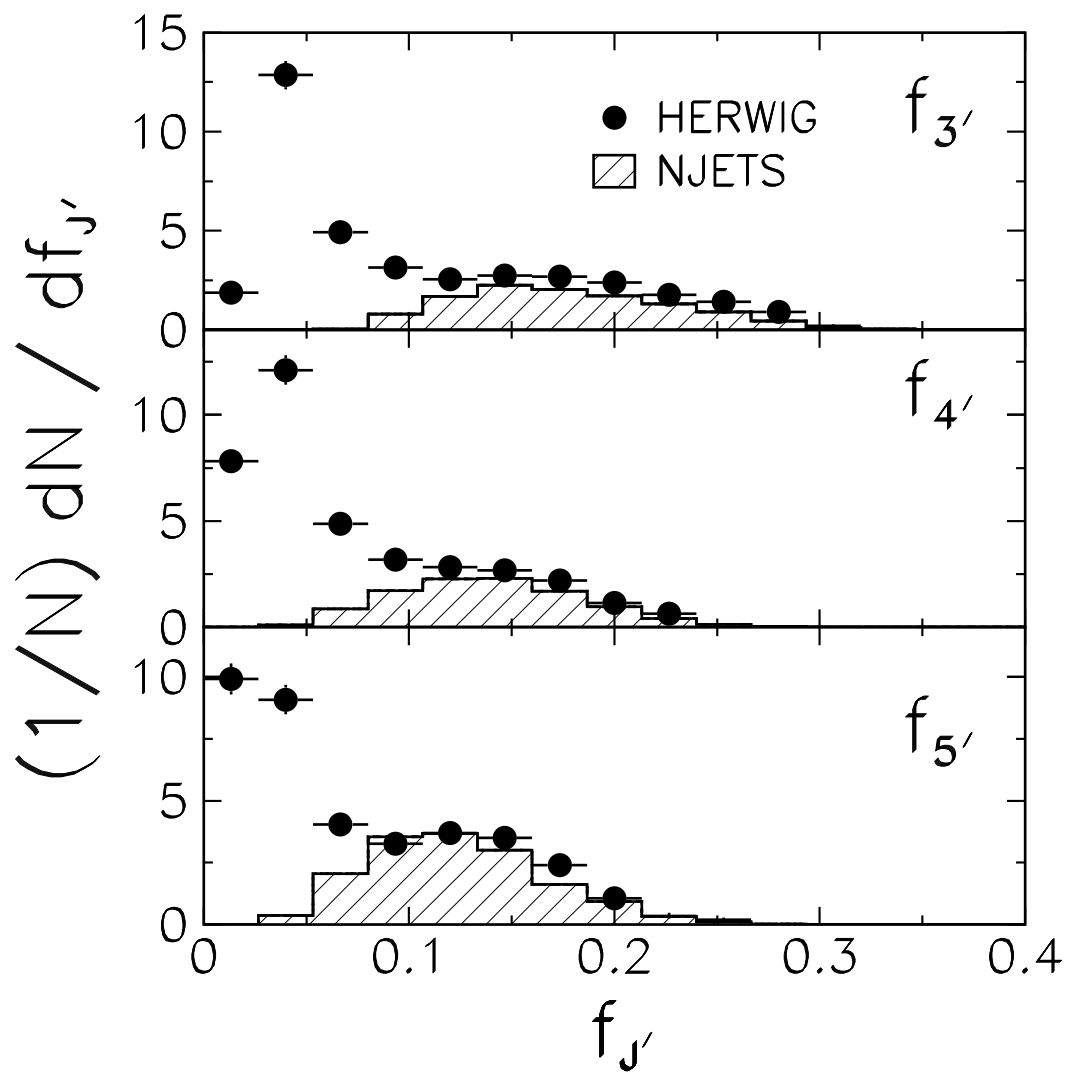}
\caption{The predicted distributions of single-jet mass-fractions for
jets in four-jet events
produced at the Fermilab Proton-Antiproton Collider that satisfy
the requirements
$m_{4J}> 650$ GeV/$c^{2}$, $X_{3'} < 0.9$,
and $\mid \cos\theta_{3'}\mid < 0.8$.
HERWIG predictions (points) are compared with NJETS predictions (histograms).}
  \label{fj_4j_fig}
\end{figure}
\clearpage
\begin{figure}
\vspace{0.5cm}
\epsfysize=3.0in
\epsffile[100 144 522 590]{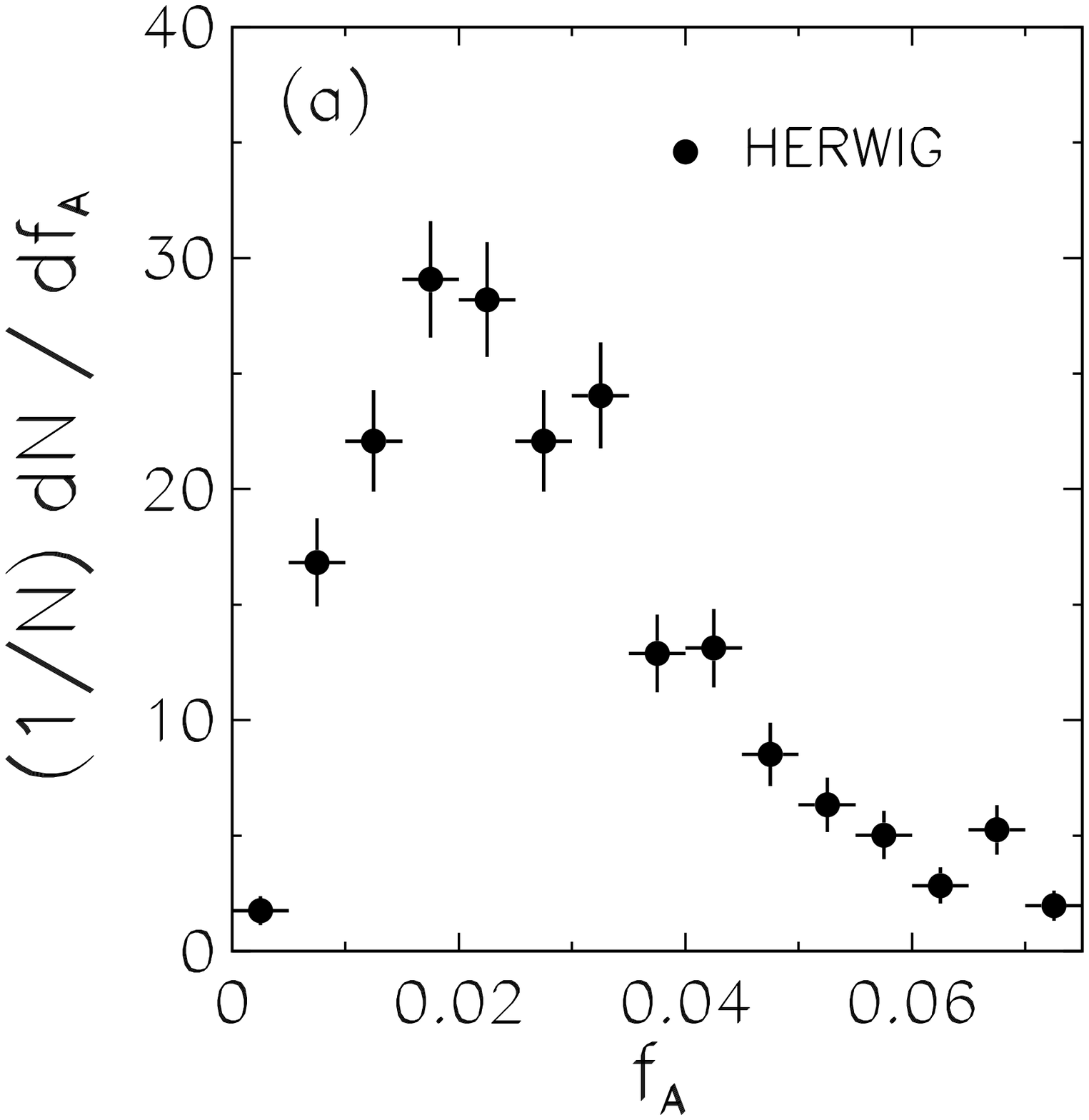}
\vspace{-3.0in}
\epsfysize=3.0in
\epsffile[-410 144 522 590]{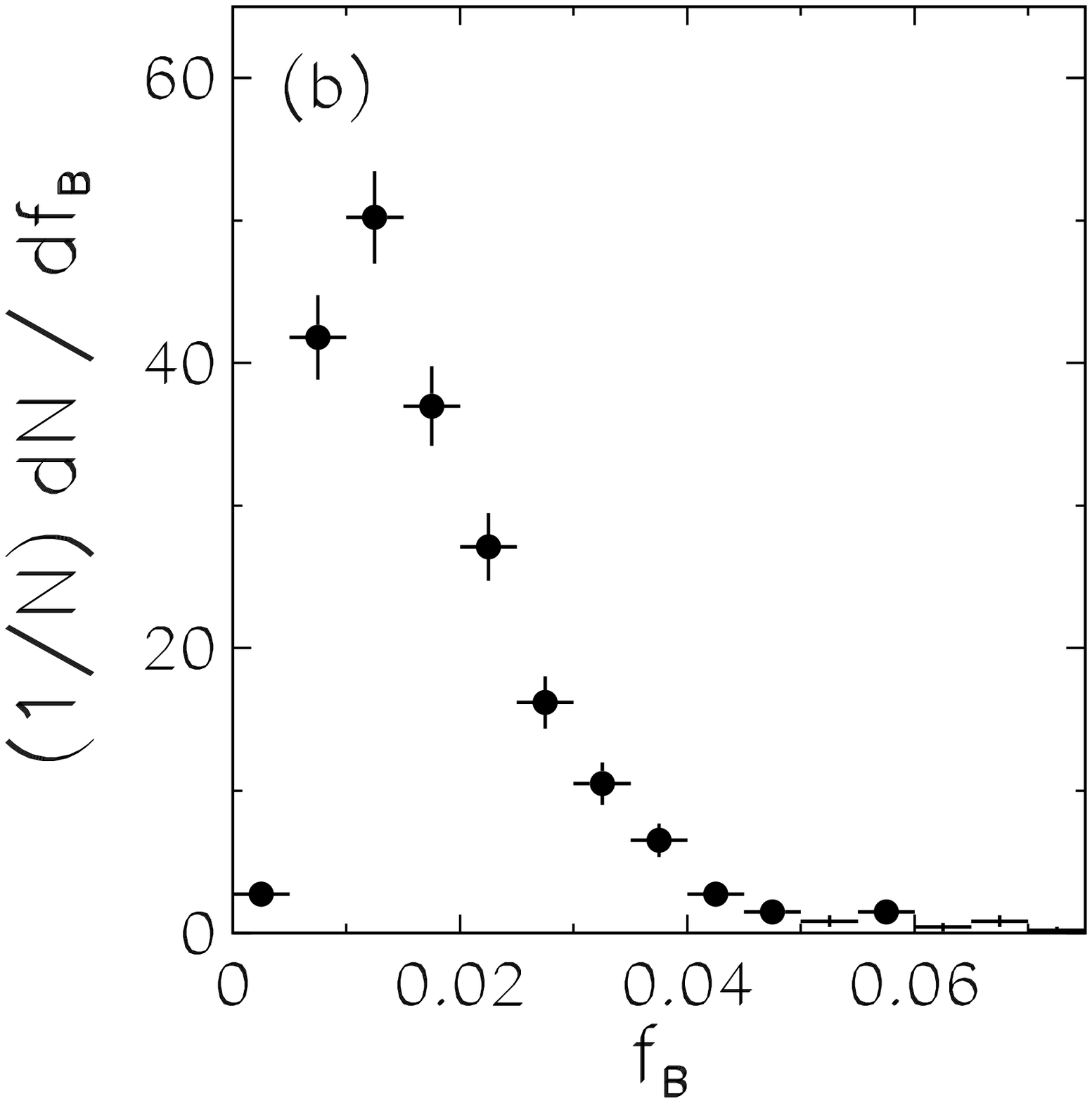}
\end{figure}
\vspace{1cm}
\begin{figure}
\epsfysize=3.0in
\epsffile[100 144 522 590]{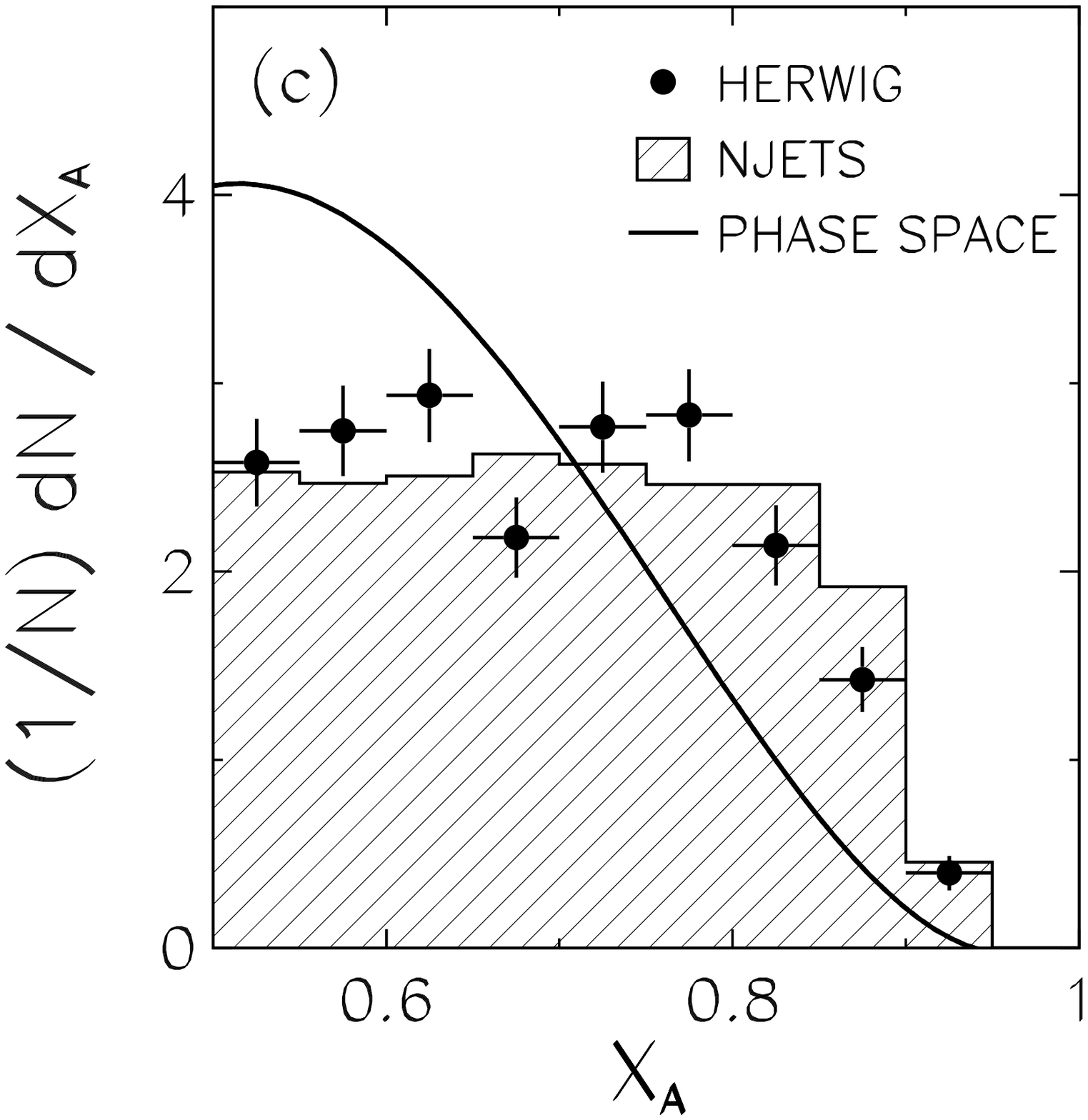}
\vspace{-3.0in}
\epsfysize=3.0in
\epsffile[-410 144 522 590]{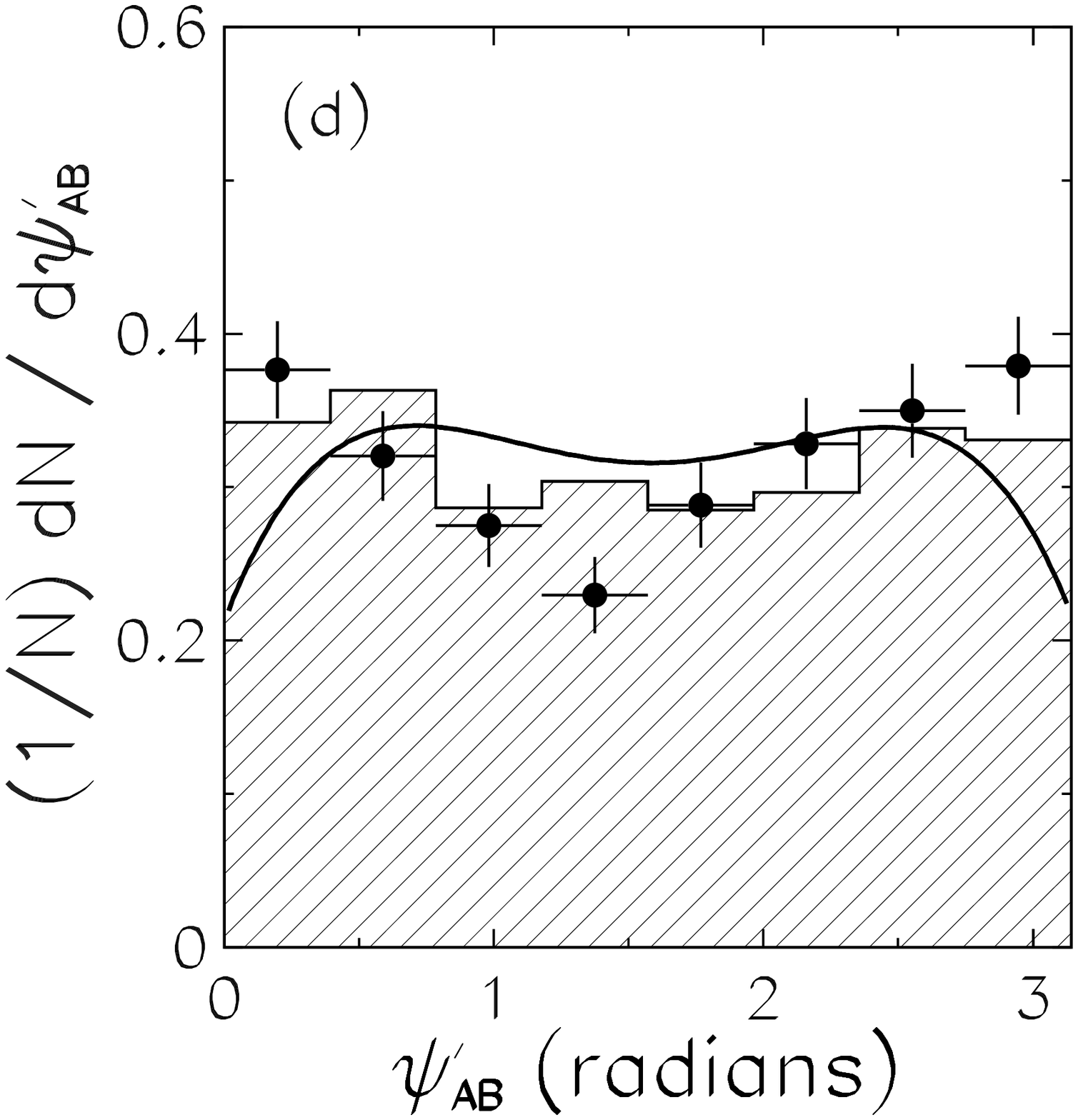}
\caption{The predicted distributions of the four-jet variables describing the
(AB)-system for four-jet events
produced at the Fermilab Proton-Antiproton Collider that satisfy
the requirements
$m_{4J}> 650$ GeV/$c^{2}$, $X_{3'} < 0.9$,
and $\mid \cos\theta_{3'}\mid < 0.8$.
The HERWIG predictions (points) are
compared with NJETS predictions (histograms), and the phase-space predictions
(curves) for
(a) $f_A$, (b) $f_B$,
(c) $X_A$, and (d) $\psi'_{AB}$.}
\label{4j_ab_fig}
\end{figure}
\clearpage
\begin{figure}
 \epsfysize=6.0in
 \vspace{-3in}
 \epsffile[-50 144 522 848]{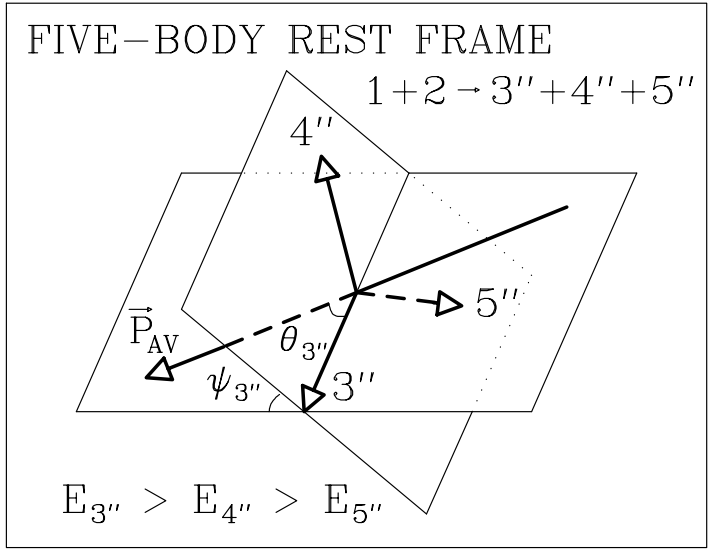}
 \epsfysize=6.0in
 \vspace{-2.5in}
 \epsffile[ -50 144 522 848]{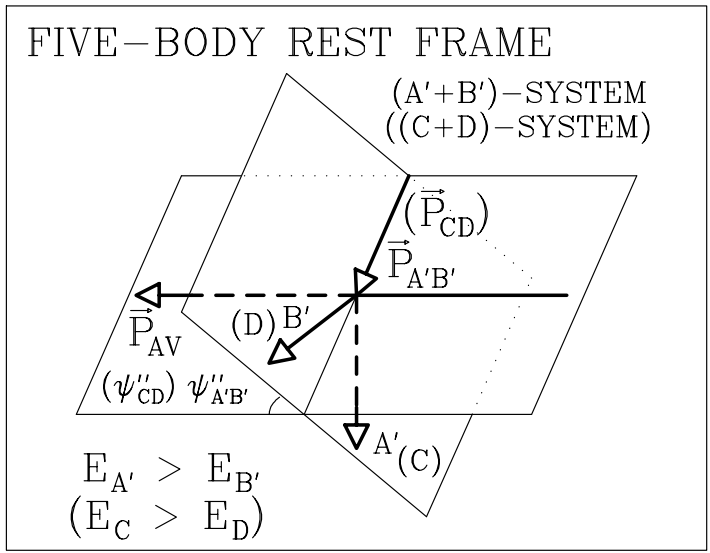}
\vspace{-1.0cm}
 \caption{Schematic definition of angles used to describe the five-jet
system in the
five-jet rest-frame.}
 \label{5jet_angle_fig}
\end{figure}
\clearpage
\begin{figure}
\epsfysize=3.0in
\vspace{0.5cm}
\epsffile[-120 144 522 590]{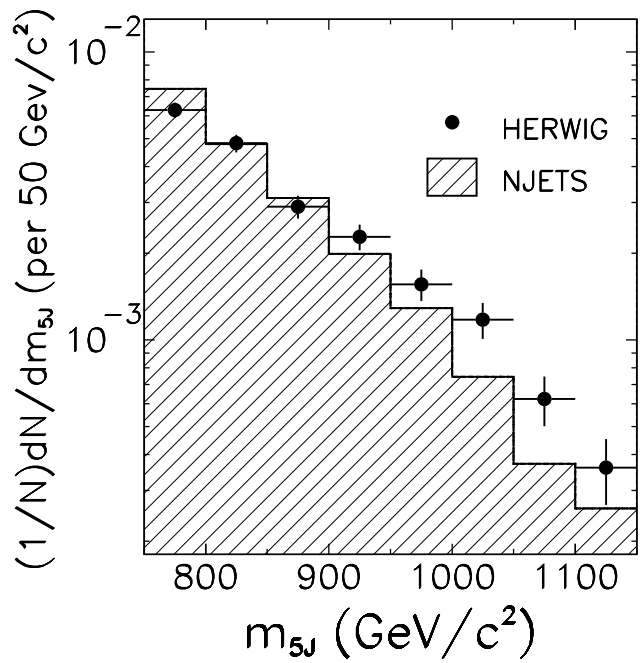}
\caption{Predicted five-jet mass distributions for five-jet events
produced at the Fermilab Proton-Antiproton Collider.
HERWIG predictions (points) compared with NJETS predictions (histogram).}
\label{m5j_fig}
\end{figure}
\clearpage
\begin{figure}
\vspace{0.5cm}
\epsfysize=3.0in
\epsffile[100 144 522 590]{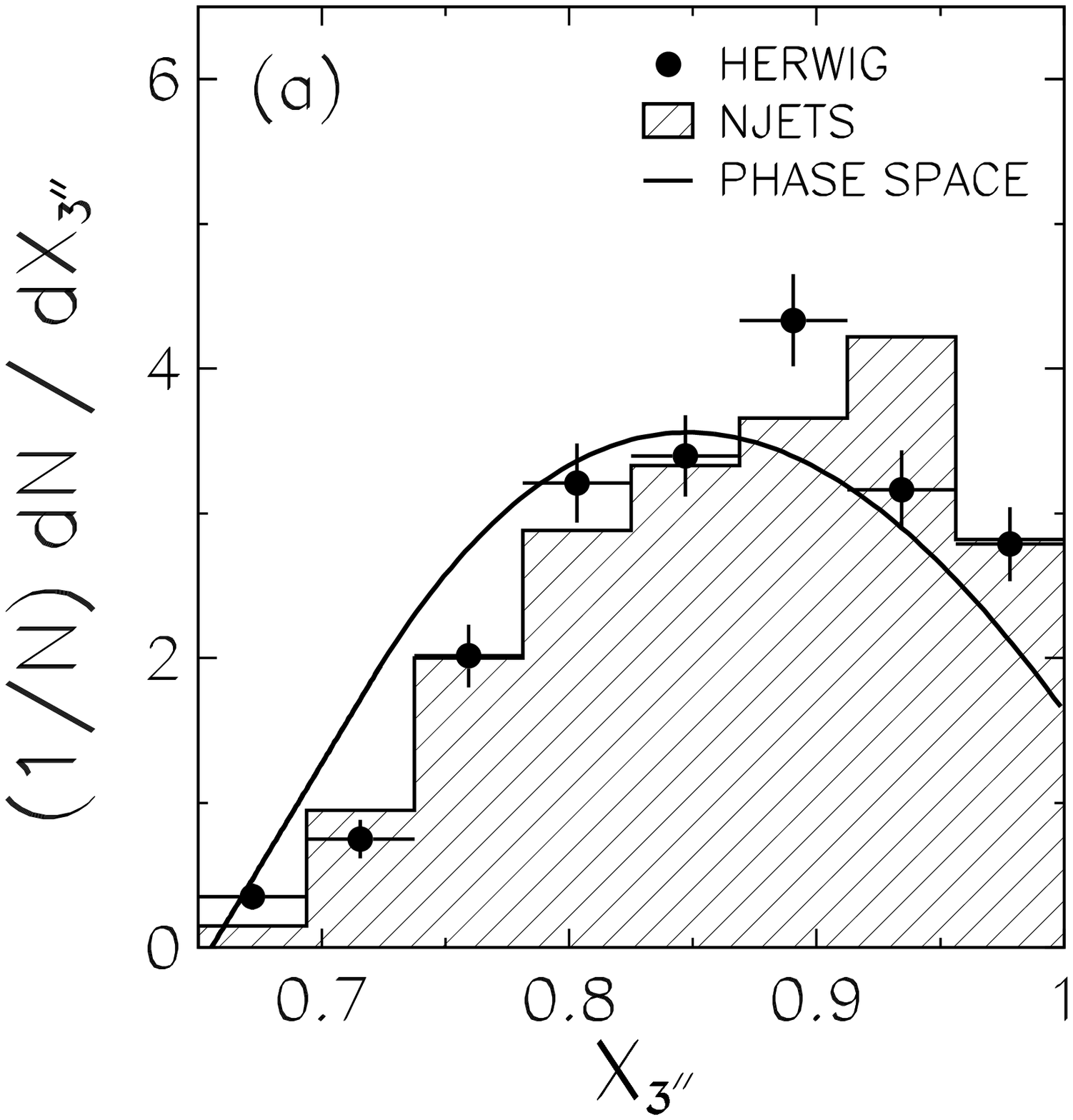}
\vspace{-3.0in}
\epsfysize=3.0in
\epsffile[-410 144 522 590]{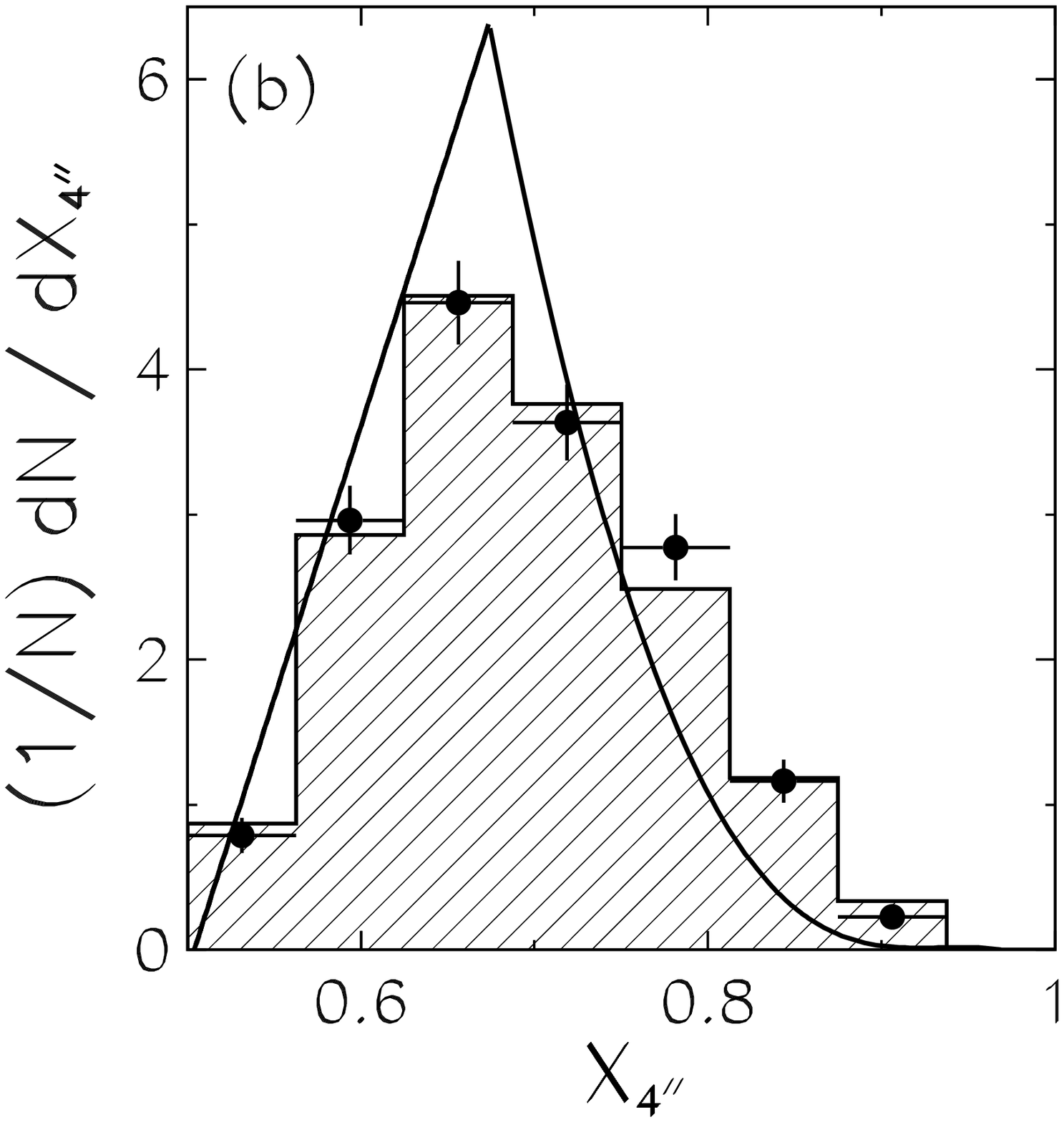}
\end{figure}
\vspace{1cm}
\begin{figure}
\epsfysize=3.0in
\epsffile[100 144 522 590]{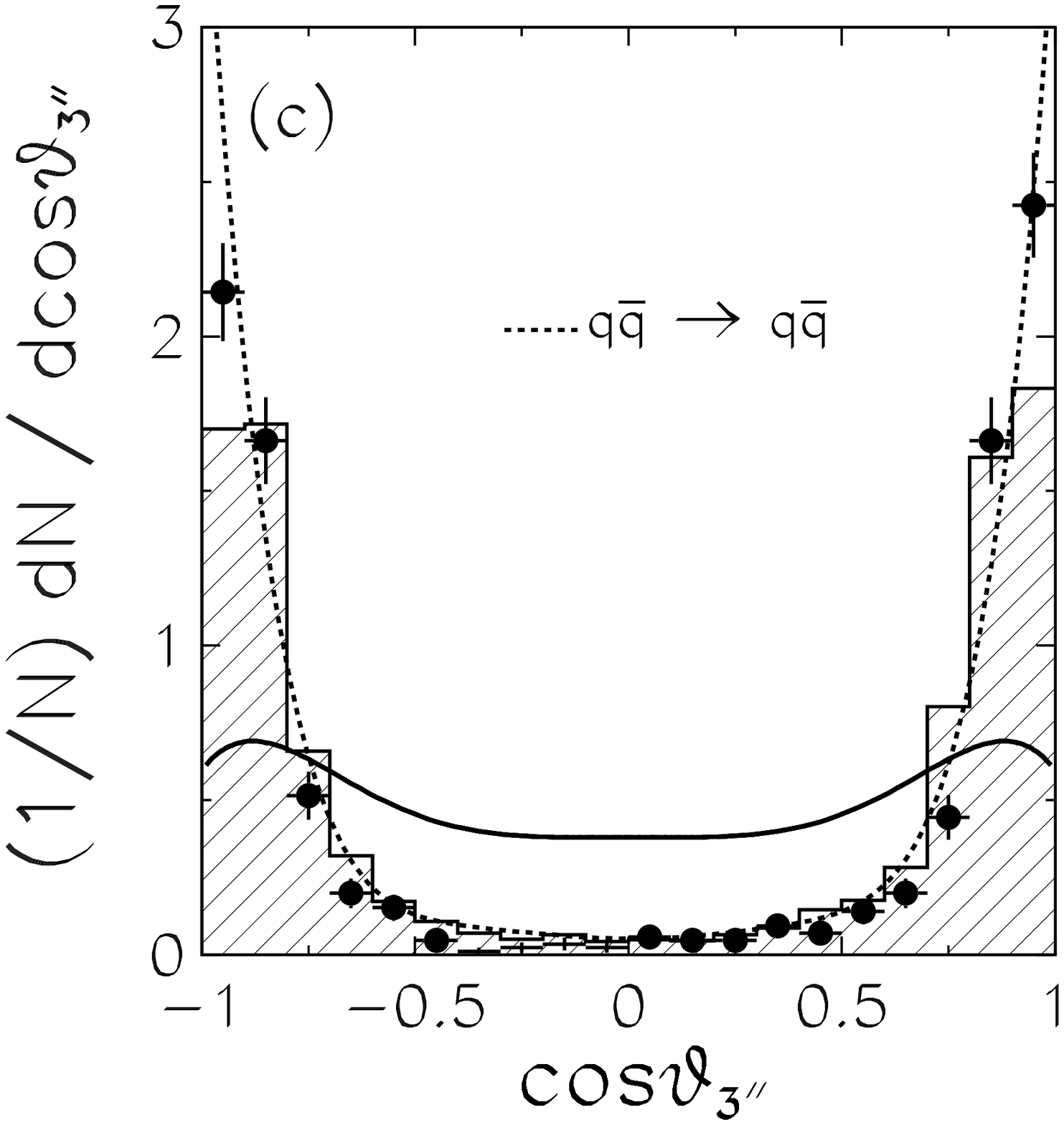}
\vspace{-3.0in}
\epsfysize=3.0in
\epsffile[-410 144 522 590]{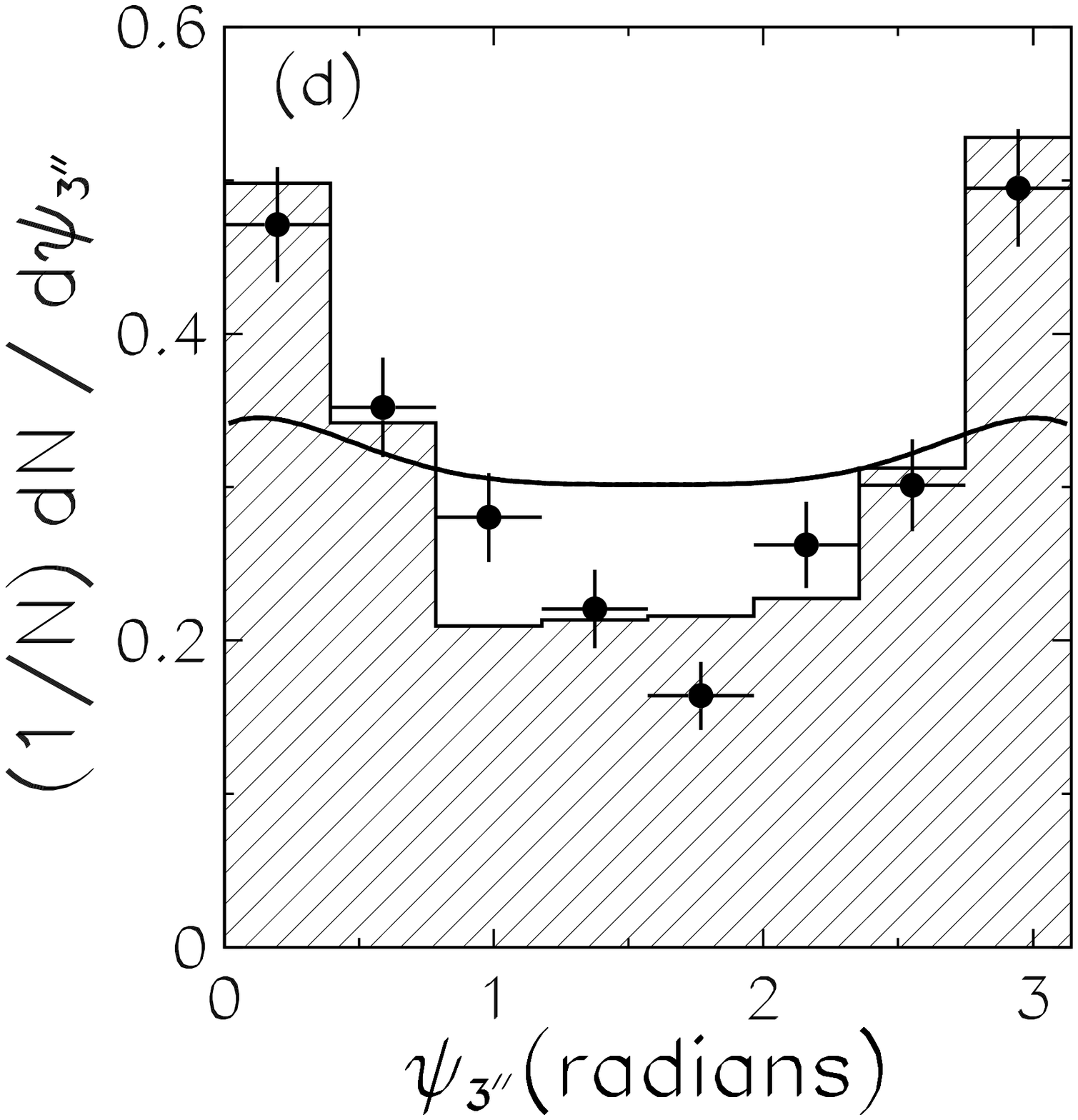}
\caption{Predicted distributions of three-body variables
for five-jet events produced at the Fermilab Proton-Antiproton Collider
that satisfy the requirement $m_{5J}> 750$~GeV/$c^{2}$.
HERWIG predictions (points) are compared with NJETS predictions
(histograms) and the phase-space predictions (solid curves) for
(a) $X_{3''}$, (b) $X_{4''}$,
(c) $\cos\theta_{3''} $, and (d) $\psi_{3''}$.
The broken curve in the $\cos \theta_{3''}$ figure is the LO QCD prediction
for $q\overline{q} \rightarrow q\overline{q}$ scattering.}
\label{5j_three_fig}
\end{figure}
\clearpage
\begin{figure}
\epsfysize=5.0in
\vspace{0.5cm}
\epsffile[30 144 522 590]{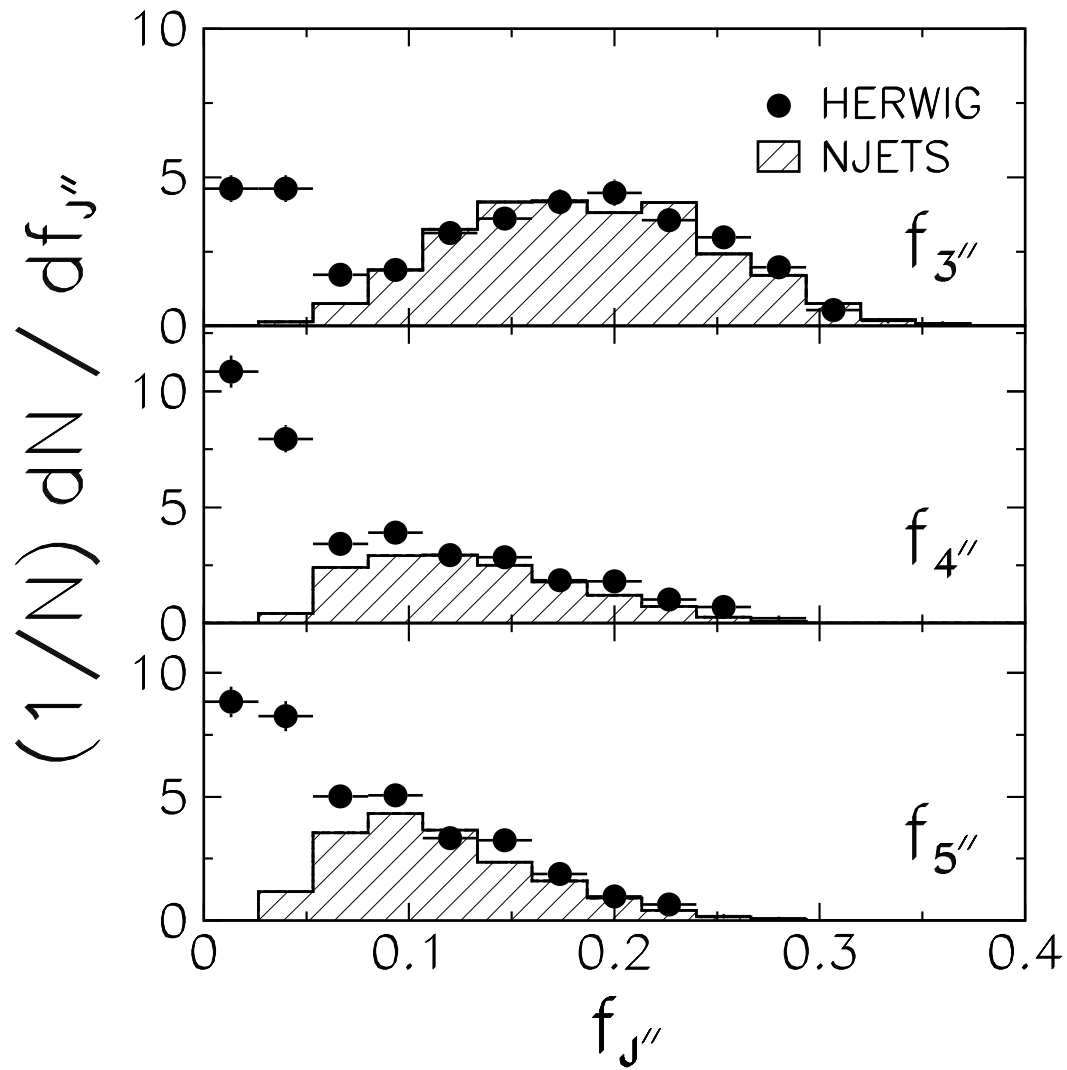}
\caption{Predicted distributions of the mass fractions described in the text
for five-jet events produced at the Fermilab Proton-Antiproton Collider that
satisfy the requirement $m_{5J}>750$~GeV/$c^{2}$.
HERWIG predictions (points) compared with
NJETS predictions (histograms).}
  \label{fj_5j_fig}
\end{figure}
\clearpage
\begin{figure}
\vspace{0.5cm}
\epsfysize=3.0in
\epsffile[100 144 522 590]{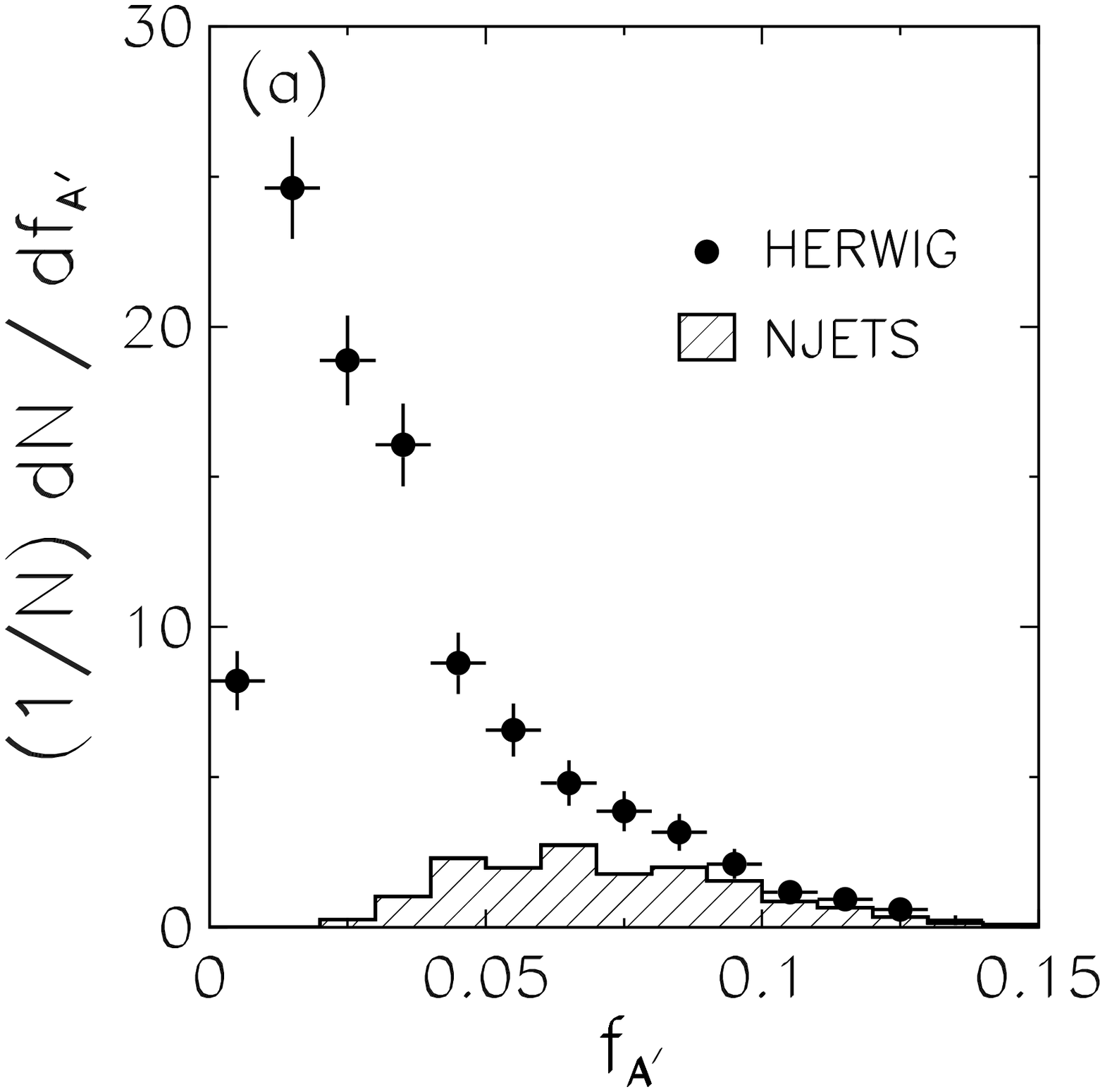}
\vspace{-3.0in}
\epsfysize=3.0in
\epsffile[-410 144 522 590]{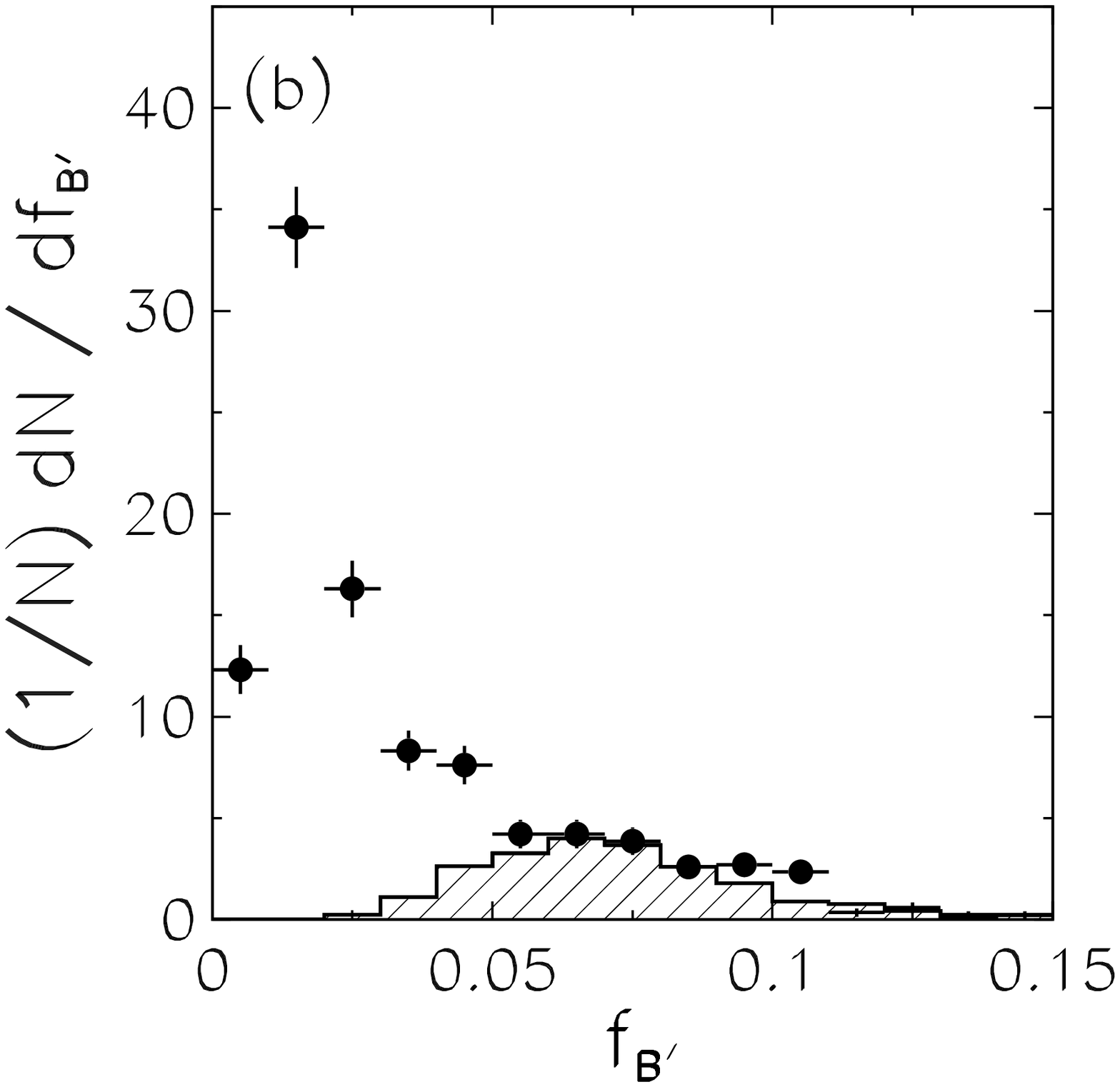}
\end{figure}
\vspace{1cm}
\begin{figure}
\epsfysize=3.0in
\epsffile[100 144 522 590]{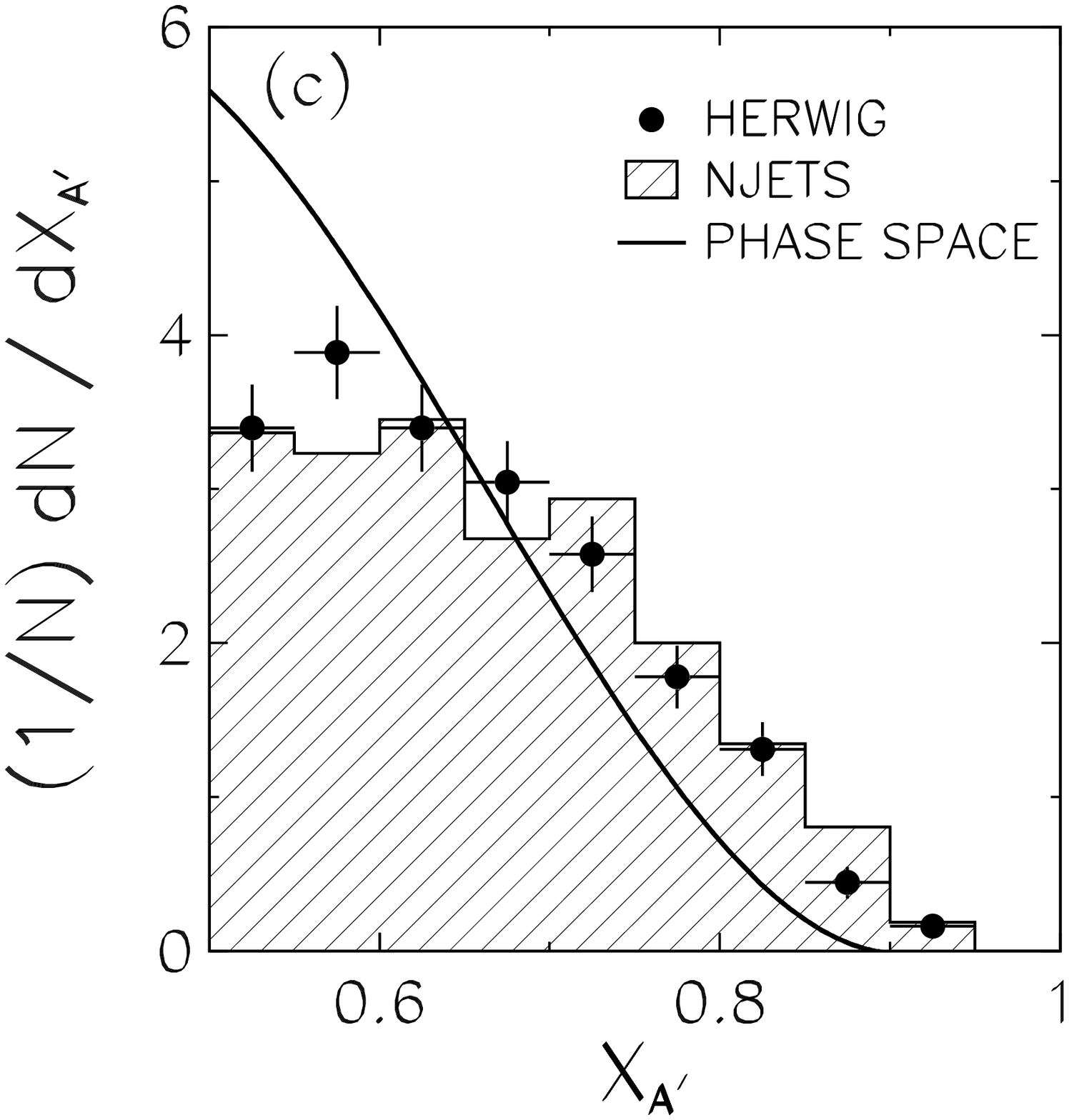}
\vspace{-3.0in}
\epsfysize=3.0in
\epsffile[-410 144 522 590]{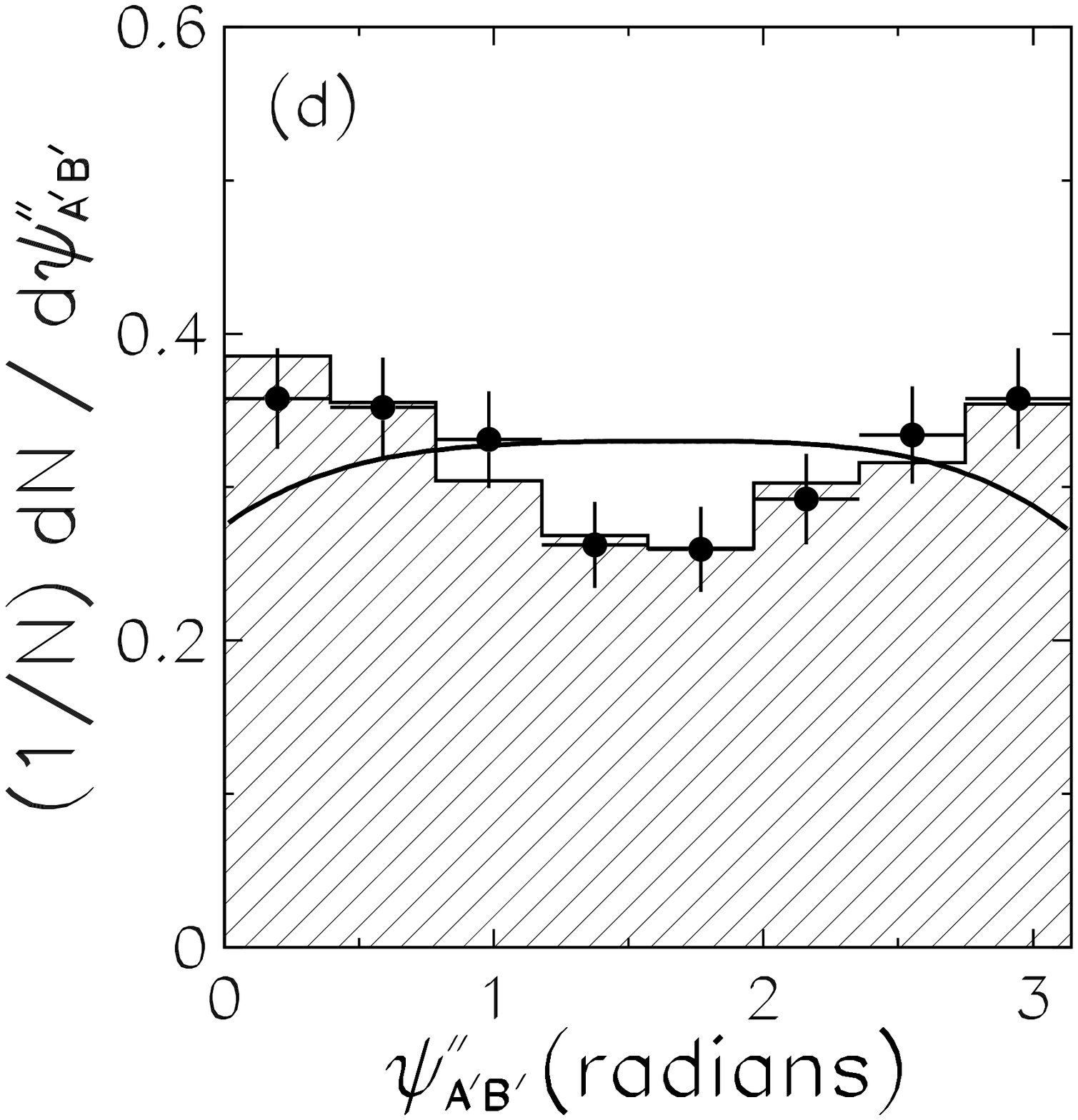}
\caption{The predicted distributions of the variables describing the
($A'B'$)-system for five-jet events
produced at the Fermilab Proton-Antiproton Collider that
satisfy the requirement $m_{5J}>750$~GeV/$c^{2}$.
HERWIG predictions (points) are compared with NJETS predictions (histograms)
and the phase-space model predictions (curves) for
(a) $f_{A'}$, (b) $f_{B'}$,
(c) $X_{A'}$, and (d) $\psi'_{A'B'}$.}
\label{5j_ab_fig}
\end{figure}
\clearpage
\begin{figure}
\vspace{0.5cm}
\epsfysize=3.0in
\epsffile[100 144 522 590]{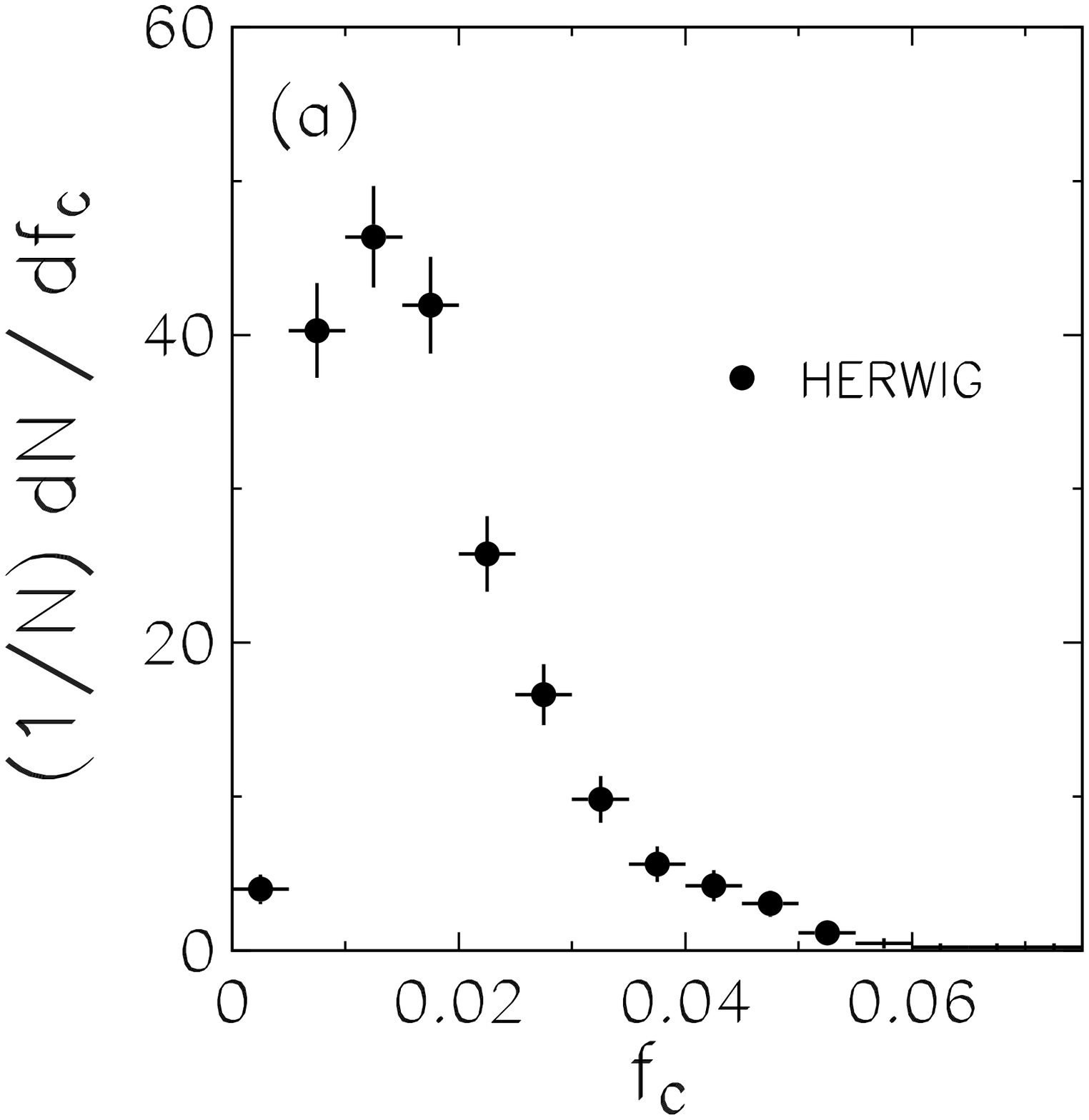}
\vspace{-3.0in}
\epsfysize=3.0in
\epsffile[-410 144 522 590]{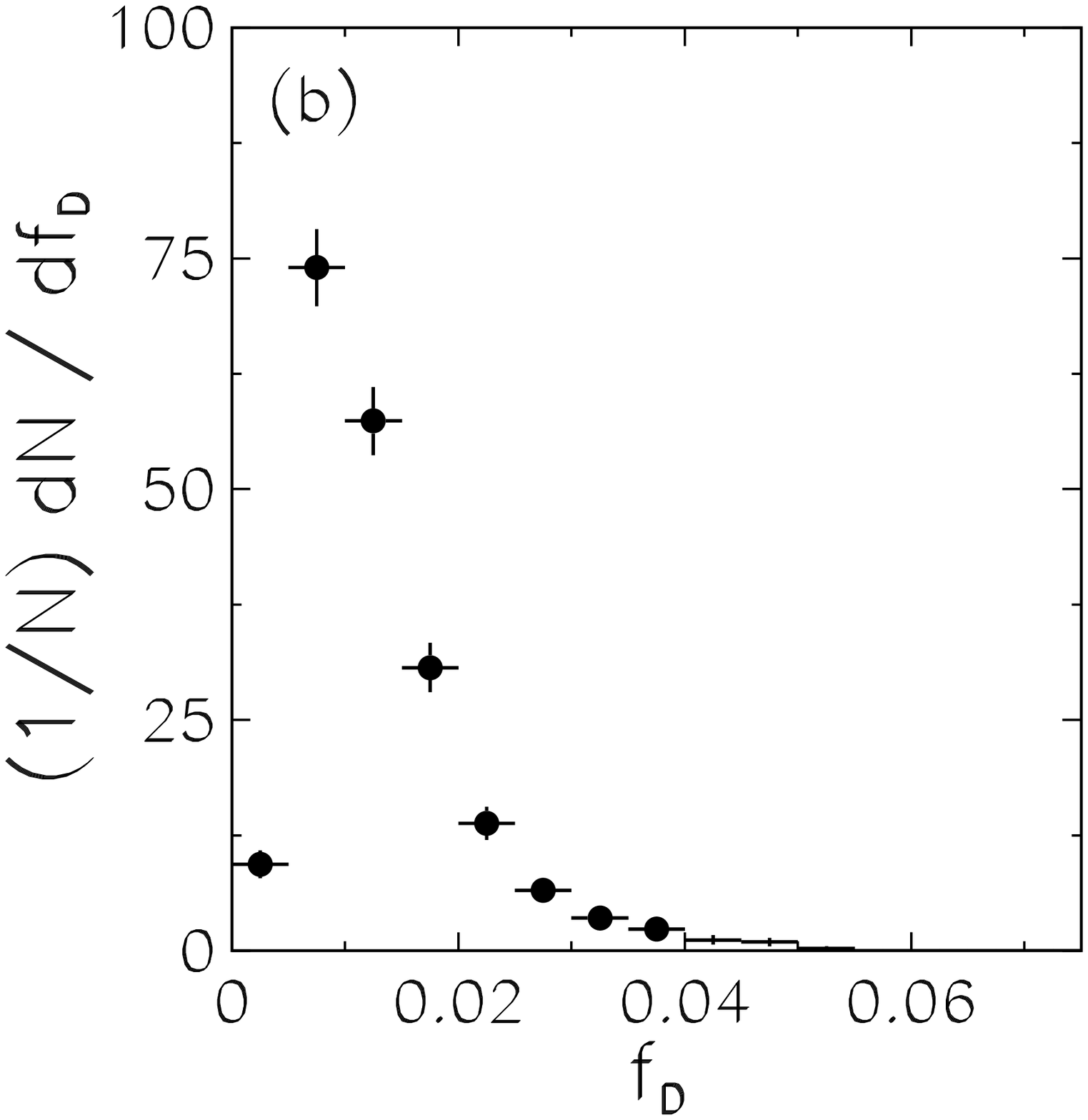}
\end{figure}
\vspace{1cm}
\begin{figure}
\epsfysize=3.0in
\epsffile[100 144 522 590]{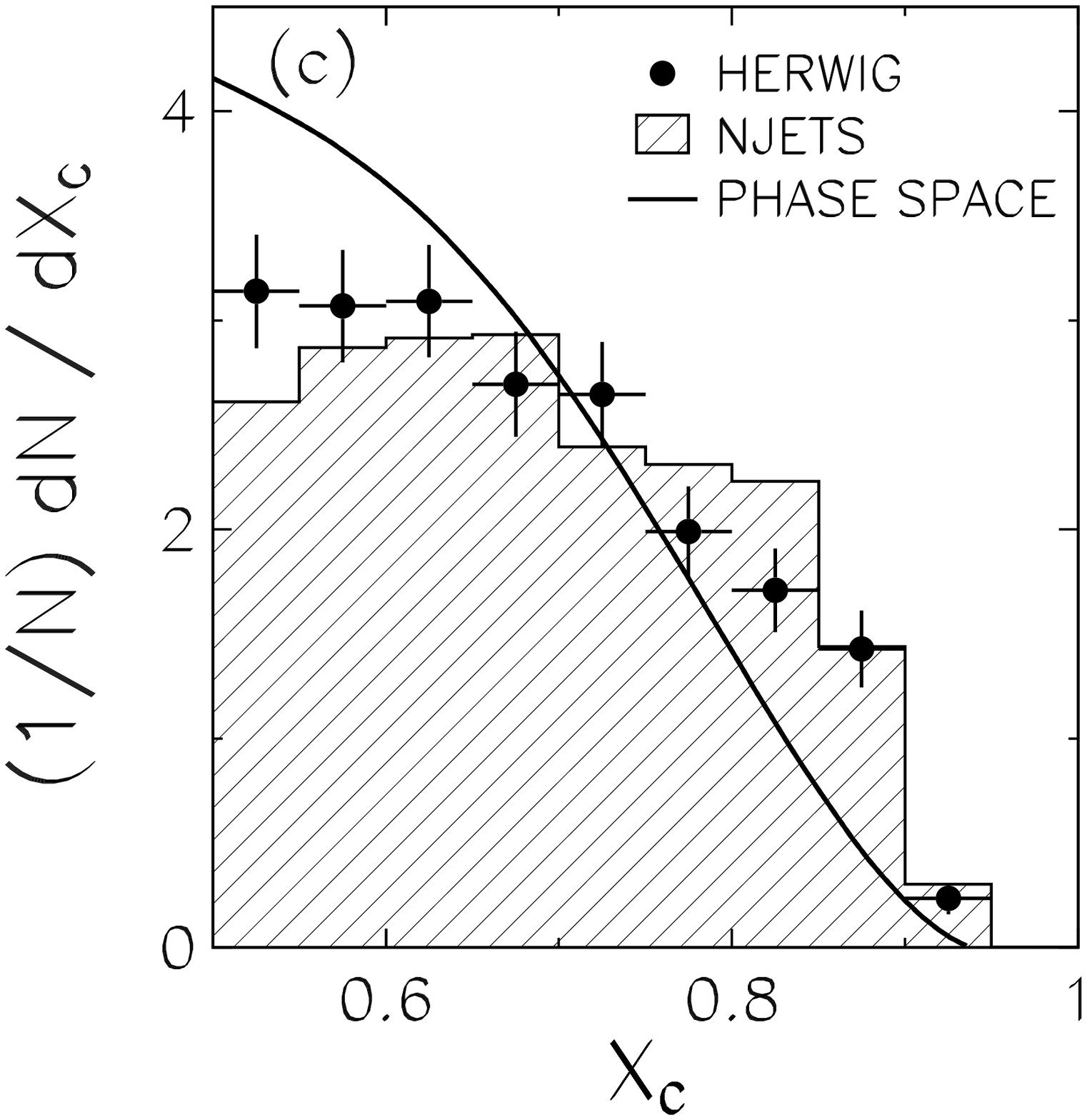}
\vspace{-3.0in}
\epsfysize=3.0in
\epsffile[-410 144 522 590]{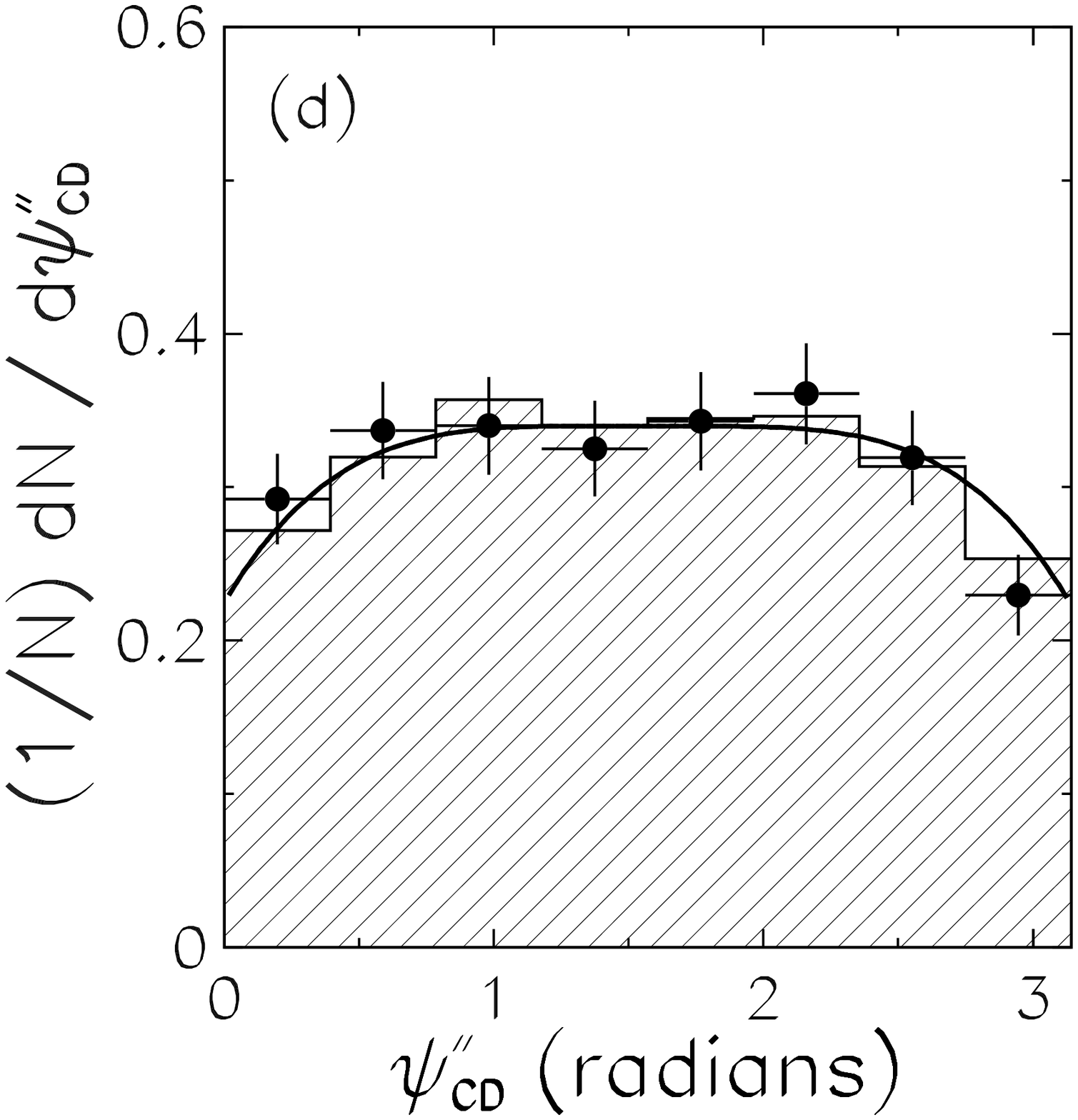}
\caption{The predicted distributions of the variables describing the
(CD)-system for five-jet events produced at the Fermilab Proton-Antiproton
Collider that
satisfy the requirement $m_{5J}>750$~GeV/$c^{2}$.
HERWIG predictions (points) are compared
with NJETS predictions (histograms) and the phase-space model predictions
(curves) for
(a) $f_C$, (b) $f_{D}$,
(c) $X_{C}$, and (d) $\psi''_{CD}$.}
\label{5j_cd_fig}
\end{figure}

\end{document}